\documentclass[12pt]{article}
\begin{document}

\title{Dynamics of FitzHugh-Nagumo excitable systems with delayed coupling}
 \author{Nikola Buri\' c  \thanks{e-mail: buric@phy.bg.ac.yu} and
Dragana Todorovi\' c   \\  Department of Physics and
Mathematics,\\ Faculty of Pharmacy, University of Beograd,\\
Vojvode Stepe 450, Beograd, Yugoslavia. }

\maketitle

\begin{abstract}

Small lattices of $N$ nearest neighbor coupled excitable FitzHugh-Nagumo
 systems, with time-delayed coupling are studied,
  and compared with systems of  FitzHugh-Nagumo oscillators with the
   same delayed coupling. Bifurcations of equilibria
 in $N=2$ case are studied analytically, and it is then numerically confirmed
 that the same bifurcations are relevant for the dynamics in the case  $N>2$.
 Bifurcations found include inverse and direct Hopf and fold limit cycle
 bifurcations. Typical dynamics for different small time-lags and
 coupling intensities could be excitable with a single
 globally stable equilibrium, asymptotic oscillatory with symmetric limit cycle,
 bi-stable with stable equilibrium and a symmetric
 limit cycle, and again
 coherent oscillatory but non-symmetric and phase-shifted. For an intermediate
 range of time-lags inverse sub-critical Hopf and fold limit cycle bifurcations
 lead to the phenomenon of oscillator death. The phenomenon does
 not occur in the case of  FitzHugh-Nagumo oscillators with the
 same type of coupling.

\end{abstract}

\newpage

\section{Introduction}

Excitability is a common property of many physical and biological
systems. Since the work of Hodgkin and Huxley \cite{HH}, and the development
 of the basic mathematical model by FitzHugh \cite{FitzHugh} and Nagumo \cite{Nagumo} the
 reported research  on the subject has grown  enormously. As for a
 general review we cite just
 the classic references \cite{excit},\cite{Winfree}
 and the references \cite{laser1},\cite{Ginsburg} for
 examples of a recent physical, and \cite{Izhikevich1}, \cite{Gerstner} for
 neuro-biological applications. For instance,
a single
 neuron displays excitable behavior, in the sense that a small perturbation away
from its quiescent state, i.e. a stable stationary value of the cross membrane potential,
 can result in a large excursion of its potential before returning to quiescent.
 Such generation of a single spike in the electrical potential across the neuron
 membrane is a typical example of the excitable behavior. Many other cells, besides
 neurons, are known to generates potential spikes across their membrane.
 Such excitable units usually appear as constitutive elements of complex systems,
 and can transmit excitations between them. The dynamics of the complex system
 depends on the properties of each of the units and on their interactions.
 In biological, as well as physical, applications the transmission of excitations
 is certainly not instantaneous, and the representation by non-local and instantaneous
 interactions should be considered only as a very crude approximation. For example,
 significant delays of more than $4\%$ of the characteristic period of
 the $40Hz$  frequency oscillations of the brain neurons,
occur during the nerve conduction between the neurons less
then $1mm$ apart \cite{Shepherd},\cite{Murray}.

This paper is devoted to an analyzes of a small lattice of
 a particular type of excitable systems, with
 a finite non-zero duration of the transfer of the excitations between the neighboring
 units. Despite its relevance and a large amount of related research ( to be summarized
 and discussed in the last section ) excitable systems with time-delayed coupling have not
 been sufficiently studied. We shall be particularly interested in the bifurcations
 that turn on and turn off the oscillatory behavior as the coupling constant and the
small time-lag are varied.

 In the introduction we formulate the model that is to be analyzed, and then briefly
preview our results and discuss
 the context of our work.

As a model of each of the excitable units we shall use
 the paradigmatic example of the FitzHugh-Nagumo system in the form, and for the
 parameter range, when the system displays the
 excitable behavior. The dynamical equations of the single uncoupled excitable
 unit are \cite{Murray}:
\begin{eqnarray}
  \dot x & = & -x^3 + (a + 1)x^2 - ax -y,\nonumber \\
  \dot y & = & bx - \gamma y.
\end{eqnarray}
where $a, b$ and $\gamma$ are positive parameters.
 In the original interpretation of (1),
 as a model of the neuronal excitability,
  $x$ represents the trans-membrane voltage and the variable $y$
  should model the time dependence of several physical quantities
   related to electrical  conductances of the relevant ion
   currents across the membrane. In the model
$x$ behaves as an excitable variable and $y$ is the slow
refractory variable.

 The particular form (1) of the
 FitzHugh-Nagumo model does not admit periodic solutions for any values of
 the parameters.
 Furthermore we shall restrict our analysis to the range of the parameter values where
 the system exhibits excitability,
 with only one attractor in the form of a stable fixed point at the origin.
 For this to be the case, $b$ and $\gamma$ should be
 of the same order of magnitude and considerably
 smaller than $a$ (see section 2).
 We refer to the system (1) in this range of parameters as the excitable
 FitzHugh-Nagumo model. On the other hand, the minimal
 modification of (1) that renders a system which could have a stable limit cycle
 is obtained by adding to the first equation an external constant
 current $I$ of a prescribed intensity. We shall refer to such a
 system with the stable limit cycle as the  FitzHugh-Nagumo
 oscillator as opposed to the  excitable
 FitzHugh-Nagumo (1).

The full system is a  one-dimensional lattice of N identical
excitable units of the form (1), given by the equations of the following type:
\begin{eqnarray}
\dot x_i & = & -x_i^3 + (a + 1)x_i^2 - ax_i -y_i+cF(x^{\tau}_{i-1},x_i,x^{\tau}_{i+1}),\nonumber \\
\dot y_i & = & bx_i - \gamma y_i,\qquad i=1,\dots N,
\end{eqnarray}
where
$$
x^{\tau}_i(t)\equiv x_i(t-\tau),
$$
and  $\tau$ is a fixed time lag and $c$ is the coupling constant. General form of the
 coupling term will be  specified later.

Local stability near the rest state of (2),
 and  global dynamics like existence of stable limit cycles, and
 the properties of the oscillations on such a cycle, do depend on the coupling function.
 However, we shall see that local properties and even the global dynamics are
 qualitatively the same for a
 large class of coupling functions, that are dependent only on the  voltages of the
 neighbors, like for example:
$$
F(x^{\tau}_{i-1},x_i,x^{\tau}_{i+1})=f(x^{\tau}_{i-1})+f(x^{\tau}_{i+1}),
\qquad f(x)=\tan^{-1}(x).
$$
On the other hand, diffusive coupling, i.e. proportional to $x_i(t)-x_{i-1}(t-\tau)$ implies
 different properties of the global dynamics.
Furthermore, important dynamical phenomena that occur for $N=2$, happen
 also for $N>2$.
In fact, most of our results will be derived by considering first the system
 with only two coupled units, and then checking the conclusions in the case of medium $N>2$
 by numerical computations.

It is well known, and often used, fact that the time-delay could destabilize a stationary
 point and introduces oscillatory behavior.
 Also, networks of oscillatory units with delayed
 coupling have been analyzed before. The studied oscillatory systems
 could be roughly divided into those where the oscillatory
 units are general limit cycle oscillators, say near the Hopf bifurcation, (for example:
\cite{Roy1},\cite{Roy2}, \cite{Wirkus}), phase-coupled phase oscillators
\cite{Schuster1},\cite{Schuster2},
\cite{Nakamura},\cite{Kim},\cite{phase2},\cite{Strogatz}), or the relaxation
 oscillators (for example: \cite{Fox},\cite{relax2})
 typical in the neuro-biological applications
\cite{Izhikevich1},\cite{Izhikevich3},\cite{Ermentrout1}.
In the later case the form of the coupling takes,
 more or less, into the account the properties of real synaptic interactions between
the neurons \cite{Ermentrout1},\cite{Ermentrout2}.

In the last section we shall more systematically compare, the system (2) and our results with
several similar or related models.
 In the introduction, we should like to point out that
 the major part of our analysis deals with the system of coupled
 excitable units, and the system of FitzHugh-Nagumo oscillators
 with the same coupling is mentioned only in order to stress the
 differences.
 On the other hand, a sufficiently strong
 instantaneous coupling (time lag equal to zero) between the excitable
 (not oscillatory) units  can introduce the oscillatory solutions.
 This phenomenon has been known already to Turing \cite{Turing} and was studied by Smale
\cite{Smale} and Johnson \cite{Jackson}. As we shall see,
for such sufficiently strong coupling, a time-lag
 which is small on the scale of the interspikes or refractory period,
 induces drastic qualitative changes in the dynamics.
 Phenomena like death of oscillations, bi-stable excitability, and transitions between symmetric
 in-phase
  and non-symmetric, phase-shifted asymptotic oscillations,
 all occur in the system (2) as the time-delay is varied. On the
 other hand, dynamics of the coupled  FitzHugh-Nagumo oscillators
 with the same type of coupling is quite different.

The results of our study are presented as follows.  Sections 2 and 3 are concerned with
 the system with just two excitable units. Analytic results about the codimension 1
 bifurcations of
 the stationary solutions are given in detail, in section 2, for a specific common type of
  coupling such that there are difference between
  coupled excitable and coupled oscillatory units. Other types of
  coupling are briefly discussed.
 Numerical analyzes of global dynamics and in particular of the periodic solutions
 and their bifurcations are presented in section 3. Here we also
 point out some of the differences between coupled  excitable systems vs. the
 oscillators.
In section 4 we demonstrate, by direct numerical computations, that the phenomena
analyzed in sections 2 and 3 for $N=2$ occur also in a similar way
 in the system consisting of $N>2$ identical units. Conclusions, discussion and
 comparison with related works are given in section 5.

\section{  Two coupled units: Local stability and bifurcations}

In this section we study stability and bifurcations of the zero stationary point of only two
 coupled identical FitzHugh-Nagumo excitable systems, given by the following equations:
\begin{eqnarray}
\dot x_1 & = & -x_1^3 + (a + 1)x_1^2 - ax_1 -y_1+cf(x^{\tau}_{2}),\nonumber \\
\dot y_1 & = & bx_1 - \gamma y_1,\nonumber \\
\dot x_2 & = & -x_2^3 + (a + 1)x_2^2 - ax_2 -y_2+cf(x^{\tau}_{1}),\nonumber \\
\dot y_2 & = & bx_2 - \gamma y_2,
\end{eqnarray}
where the coupling function satisfies
\begin{equation}
f_0=0,\qquad\ f'_0=\delta > 0,
\end{equation}
and the subscript $0$ denotes that the function is evaluated at
$(x_1,x_2,y_1,y_2)=(0,0,0,0)$. If fact, the first condition is not
crucial, and is
 introduced only for convenience.

{\it A single neuron}:

Consider first one of the units in the case of zero coupling. Point $(x,y)=(0,0)$ is
 an intersection of the qubic $\dot x$ nullcline
 and the linear $\dot y$ nullcline for any value of the parameters $a,b,\gamma$, so that
 it is always a stationary point. Furthermore it is always a stable stationary
 point, that could be a node, if $(a-\gamma)>2\sqrt b$,  or a focus,
 when $(a-\gamma)<2\sqrt b$.
There could be one more (non-generic case) and two more
stationary points, but we shall restrict our attention to the case when $(0,0)$ is the
 only stationary solution. This is the case if $4b/\gamma>(a-1)^2$.
 We shall make no further assumptions as to the nature of
 the stable stationary point $(0,0)$, but, as we shall see, some of the typical
 behavior of the delayed coupled systems will be lost in the singular limit
 $b\rightarrow 0,\>\gamma\rightarrow 0$, and difficult to observe very close to
 this limit.
 The particular form of the FitzHugh-Nagumo model, with no external current
 and $(0,0)$ stationary point does not possess periodic solutions for any values
 of the parameters. However,  there are
  solutions that start in a small neighborhood of $(0,0)$, quite rapidly go relatively far
 away, and then approach back on to the stationary point,
 (see figure 5a). Such solutions represents typical excitable
 behavior.  The excitability that is displayed by the FitzHugh-Nagumo system is of,
 the so called, type II \cite{Izhikevich1}, in the sense that there is no clear-cut
 threshold in the
 phase space between the excitable orbits and the orbits that return quickly,
 and directly, to the rest state.
 In fact, there are orbits which
 continuously interpolate between the two types of orbits.
 However, as the parameters $b$ and $g$ are decreased, as compared with a
 fixed $a$,
 the excitable behaviour quite rapidly (but continuously) become dominant.
  We shall always use such values of the parameters that the
  excitable behaviour is clearly demonstrated, for example $b>\gamma^2$
  and $a\gg b,\> a\gg\gamma$.

  In order to stress typical properties of the excitable, but not oscillatory,  systems,
  we shall also need a convenient system with a stable oscillatory
  behaviour. Such a system is obtained by adding an external (say, constant)
  current $I$ to the $\dot x$ equation (or to the other one) of the FitzHugh-Nagumo
  model. The constant current shifts the intersection of the two
  nullclines, and if it is such that the intersection lies on the
  part of the $\dot x$ nullcline with a positive slope, then the
  stationary point is unstable and there is a stable limit cycle.
  The limit cycle is born in the supercritical Hopf bifurcation of
  the stationary solution.  The limit cycle is of approximately circular
  shape only if $I$ is quite close to the critical value $I_0$,
  and then the amplitude is of the order of $\sqrt{I-I_0}$.
   Otherwise it has the shape typical for relaxation oscillators.

{\it Instantaneously coupled identical units}:

As the next step, let us fix the parameters $a,b$ and $\gamma$ such that each of the units
 displays the excitable behavior, and consider the coupled system but with the
 instantaneous coupling $\tau=0$. Point $(x_1,y_1,x_2,y_2)=(0,0,0,0)$ represent a stationary
 solution, and its local stability is determined by analyzing the corresponding
 characteristic equation
\begin{equation}
[( a+\lambda)(\gamma+\lambda)+b-c\delta(\gamma+\lambda)]
[( a+\lambda)(\gamma+\lambda)+b-c\delta(\gamma+\lambda)]=0,
\end{equation}

The sign of the real parts of the four eigenvalues
\begin{eqnarray}
2\lambda_{1,2} & = & -( a+\gamma-c\delta)\pm \sqrt {( a-\gamma-c\delta)^2-4b},\nonumber \\
2\lambda_{3,4} & = & -( a+\gamma+c\delta)\pm \sqrt {( a-\gamma+c\delta)^2-4b},
\end{eqnarray}
determines the stability type of the trivial stationary point. If $(a-\gamma)>2\sqrt b$ the
 point is stable
 node-node for $0<c<(a-\gamma-2\sqrt b)/\delta$, and if $c$ is larger the eigenvalue
 $\lambda_{1,2}$ becomes complex, and the point becomes stable focus-node. Otherwise, if
 $(a-\gamma)<2\sqrt b$ the point is stable focus-focus for $0<c<(2\sqrt b-a+\gamma)/\delta$
 and for larger $c$ the eigenvalue $\lambda_{3,4}$ becomes real and the point is
 again stable focus-node.

 Thus, whatever the stability type of the stationary point
 in the uncoupled case might
 be, there is the corresponding value of the coupling constant $c$ such that the point
 becomes focus-node.  Then, the complex pair of the eigenvalues  $\lambda_{1,2}$
 correspond to the eigenspace given by $x_1=x_2$ and $y_1=y_2$.
 In such a situation the damped oscillations of the
 two units  interfere synchronously, and  at
 some still larger $c_0$ given, in both the cases, by
\begin{equation}
c_0={a+\gamma\over \delta}
\end{equation}
where
\begin{equation}
sgn\left({d{\rm Re} \lambda_{1,2}\over dc}\right)_{c=c_0}=sgn\left(\delta/2\right)>0,
\end{equation}
 the point goes through a direct supercritical Hopf bifurcation.
 As the result, for small $\epsilon=c-c_0>0$,
 the stationary solution acquires an
 unstable two-dimensional manifold, with a stable limit cycle in it.
 The unstable
 manifold is in fact a plane given by the equations $x_1=x_2\equiv x$ and $y_1=y_2\equiv y$,
 independently of the form of the coupling function as in (3).
 Oscillations on the limit cycle are synchronous, with the linear frequency
 $\omega=\sqrt{b-\gamma^2}$, and symmetrical.
In this paper, by synchronous we mean coherent in-phase oscillations,
 and by symmetrical we mean that $x_1(t)=x_2(t)$.

The dynamics on the unstable manifold
 for small $\epsilon$ is given by the normal form of the Hopf bifurcation
 \begin{equation}
\dot r=d\epsilon r+\alpha r^3+O(\epsilon^2 r,\epsilon r^3,r^5),\qquad
\dot \theta=\omega+e\epsilon+\beta r^2+O(\epsilon^2,\epsilon r^2,r^4),
\end{equation}
where $\omega=\sqrt{b-\gamma^2}$, $d=\delta/2$, $e=-\gamma\delta/2\omega$
 and $r$ and $\theta$ are the polar coordinates
$$
x=r\sin\theta,\qquad y=r\cos\theta.
$$
  The parameters $\alpha$ and $\beta$  depend on the particular
 form of the coupling function.
For example, in the case the coupling function is
$f(x)=\tan^{-1}(x)$, then
$$
\alpha={-3+c_0\over 8}+{(a+1)^2\gamma\over 4\omega^2},
\quad \beta={(3+c_0)\gamma\over 8\omega}-{(a+1)^2(5\gamma^2+2\omega^2)\over 12\omega^3}
$$

 The limit cycle of (9) is a good
approximation only for $\epsilon$ quite small. However numerical
analysis shows that the limit cycle remains a global attractor in
the full four-dimensional
 phase space of the system (3) with no time-delay
 for a large range of $c>c_0$ values, where the approximation by the Hopf normal form (9)
 is no more valid.
  Thus there is a range of values of the coupling
 parameter $c$ where the system (3) with no time-delay
 behaves as a system of two coupled identical limit cycle oscillators.
 Properties of the oscillations on the limit cycle depend on $c$.
 Perturbation analyzes for $\epsilon$ small, or the numerical analyzes for larger $c$,
 show that oscillations on the limit cycle are synchronous and symmetrical.
 In figure 1, we illustrate the limit cycles in the coupled
excitable systems with no time-delay.
 The figure illustrates oscillatory dynamics of both units since on the limit cycle
 $x_1(t)=x_2(t)$ and $y_1(t)=y_2(t)$.
Although the  limit cycles deform continuously with
 $c$, the deformation from the small Hopf circle all the way up to the large limit cycle
 of the shape like for the relaxation oscillators, happens on a small interval of the
 values of $c$, smaller than $3 \%$ of the interval $(c_0,c_1)$.

  Further increase of $c$, still with $\tau=0$, leads to a bifurcation of
the stationary solution
 and of the limit cycle. For $c>(a-\gamma+2\sqrt b)/\delta$,
 there is a pair of real positive and a pair of real-negative eigenvalues at
 the trivial solution.  The limit cycle disappears at some still larger $c_1$
 when there appear other stable stationary solutions of (3) (with $\tau=0$).
 This value of the coupling constant $c=c_1$, when there appears nonzero stable stationary
 solution, obviously depends on the coupling function.

 In conclusion, there are three qualitatively different types of
 dynamics of the instantaneously coupled excitable systems. For
 $0<c<c_0$ the coupled system behaves as a pair of excitable
 units, while for $c_0<c<c_1$ the system behaves as a pair of identical
 limit cycle oscillators. For $c>c_1$ there appears a nontrivial
 stable stationary state.
 However, we shall be interested in the influence
 of time-delay only when the coupling constant is in the range
 $c\in (0,c_1)$, i.e. when the instantaneously coupled system behaves
 either as excitable $c<c_0$ or as oscillatory $c>c_0$.

Let us now briefly consider the coupled FitzHugh-Nagumo
oscillators, just in order to stress the
 properties which are relevant for comparison of the influence of the time-delay on the
dynamics of coupled oscillatory vs. excitable  FitzHugh-Nagumo
systems. Thus the external current $I\neq 0$ and in the range such
that each of the non-coupled units is an oscillator, either close
to the Hopf bifurcation or of the relaxation type.  We consider
the coupling of the same type like in the case of the coupled
excitable units (3) and (4).
 For convenience,  the zero of the
coupling function is shifted to coincide with the unstable
stationary point of the non-interacting oscillators.
 The major effect of such coupling is to
increase the amplitude of each of the  oscillators. The amplitude
monotonically increases with the coupling constant $c$.
 Furthermore, for
 the positive coupling constant smaller then some value the
 asymptotic dynamics of the oscillators is symmetric.
 However, the oscillations of the two units on the attractor need not be
 in-phase
  for larger values of the coupling constant, contrary to the case with
 oscillations in the instantaneously coupled excitable systems.

 Before we pass onto the analysis of the delayed equations, let us mention that
 the diffusive coupling, when for in the $\dot x_1$ and $\dot x_2$ equations one has
 $(x_1-x_2)$ and $(x_2-x_1)$ respectively, also leads to
 destabilization of the stationary point and appearance of the stable limit cycle.
 However, in this case $x_1(t)\neq x_2(t)$ and the corresponding oscillations are
 coherent but with a constant phase lag.  On the other hand, the
 trivial stationary point of the system with reversed
 diffusive coupling is stable for any positive $c$, even with an arbitrary time lag.

{\it Delayed coupling}:

Let us now consider the dynamics in a neighbourhood of the
stationary point of the delayed system (3). The point
$(x_1,y_1,x_2,y_2)=(0,0,0,0)$  is also the stationary solution of
(3), but its stability depends on $\tau$. Linearization of the
system and  substitution $x_i(t) = A_ie^{\lambda t}$, $y_i(t) =
B_ie^{\lambda t}$, $x_i(t - \tau) = A_ie^{\lambda(t - \tau)}$,
results in a system of algebraic equations for the constants $A_i$
and $B_i$.
 This system has a nontrivial solution if the following is
 satisfied:
\begin {equation}
\Delta_1(\lambda) \Delta_2(\lambda)=0,
\end{equation}
where
\begin{eqnarray}
 \Delta_1(\lambda)&=&[\lambda^2+( a+\gamma)\lambda+ a\gamma+b-c\delta\lambda\exp(-\lambda\tau)
-c\delta\gamma\exp(-\lambda\tau)]\\
\Delta_2(\lambda)&=&[\lambda^2+( a+\gamma)\lambda+ a\gamma+b+c\delta\lambda\exp(-\lambda\tau)
+c\delta\gamma\exp(-\lambda\tau)]
\end{eqnarray}

 The equation (10) is the characteristic equation of the system (3).
  Infinite dimensionality of the system is reflected in
 the transcendental character of (10). However, the spectrum  of the
 linearization of the equations (3) is discrete and can be divided into
 infinite dimensional hyperbolic and finite dimensional
 non-hyperbolic parts \cite{Lunel}.
  As in the finite dimensional
 case, the stability of the stationary point is
 typically, i.e.
  in the hyperbolic case, determined by the signs of the real parts of the roots
  of (10). Exceptional roots, equal to zero or with zero real part,
  correspond to the finite dimensional center manifold where the
  qualitative features of the dynamics, such as local stability, depend on the nonlinear terms.

 Let us first answer the question of local stability of the
 stationary point for all time-lags. We have proved (see Appendix 1) that
  the
 stationary point remains locally stable for all time-lags if the
  coupling constant is below some value $c^{\tau}$, which is smaller
  than $c_0$, given by

 \begin{equation}
  c^{\tau}=\sqrt{ {a^2\gamma^2-2b+2\sqrt{
  b(2\gamma^2+2a\gamma+b^2)}\over \delta^2}}<c_0
  \end{equation}
The previous expression for $c^{\tau}$ is valid if $b>\gamma^2$
which is always satisfied by initial assumptions about the
excitable units. Notice that the interval $(c^{\tau},c_0)$ is
quite small for the units that display excitable behaviour, and
shrinks to zero length as $(b+\gamma)/a\rightarrow 0$.

  Thus there could be two qualitatively different types of local
  dynamics around the stationary solution of the time-delayed
  coupled excitable ($c<c_0$) units. The stationary solution could
  be a combination of the stable node or the stable focus for $c<c^{\tau}$ and any
  $\tau$, and for $c^{\tau}<c<c_0$ and sufficiently small
  time-lags, or it could be an unstable focus  for $c>c^{\tau}$ and
  for some  time-lags larger than a critical value. The smallest
  critical time-lag will be found by studying the bifurcations
  conditions. In the next section, we shall see that there is also
  an important global bifurcation due to sufficiently large $\tau$ inside the interval $(0,c_0)$
  which changes the dynamics of the excitations.

  Bifurcations due to a non-zero time-lag occur when some of the roots of (10) cross the imaginary axes.
Let us first discuss the nonzero pure imaginary roots.
 Substitution $\lambda=i\omega$, where $\omega$ is real and positive, into the
 first factor gives
\begin{eqnarray}
 c\delta (\omega^2+\gamma^2)\sin(\omega\tau)&=& -\omega^3+(b-\gamma^2)\omega
\nonumber\\ c\delta (\omega^2+\gamma^2)\cos(\omega\tau)&=&
a\omega^2+( a\gamma+b)\gamma \nonumber
\end{eqnarray}
 or into the second factor
\begin{eqnarray}
 c\delta (\omega^2+\gamma^2)\sin(\omega\tau)&=& \omega^3-(b-\gamma^2)\omega
\nonumber\\ c\delta
(\omega^2+\gamma^2)\cos(\omega\tau)&=&-a\omega^2-(
a\gamma+b)\gamma\nonumber
\end{eqnarray}
Squaring and adding the previous two pairs of equations  results
in the same equation
\begin{equation}
\omega^6+(A+\gamma^2)\omega^4+(\gamma^2A+B)\omega^2+b\gamma^2=0
\end{equation}
where
\begin{equation}
  A= a^2+\gamma^2-2b-c^2\delta^2,\quad {\rm and}\quad B=(
a\gamma+b)^2-c^2\delta^2\gamma^2.
\end{equation}

Since $\omega^2\neq -\gamma^2$ the term $\omega^2+\gamma^2$ can be
factored out from (14) to obtain
\begin{equation}
\omega^4+A\omega^2+B=0.
\end{equation}

 Solutions of (16) give the frequencies $\omega_{\pm}$ of possible non-hyperbolic solutions
\begin{equation}
\omega_{\pm}=\sqrt{(-A\pm\sqrt{A^2-4B})/2}
\end{equation}
The corresponding critical time lag follows from equations (13) and (14).
 Consider the first set (13). Then, if
\begin{equation}
\sin(\omega\tau)={-\omega_{\pm}^3+(b-\gamma^2)\omega_{\pm}
\over c\delta(\omega^2_{\pm}+\gamma^2)}>0
\end{equation}
we have
\begin{equation}
\tau^j_{1,\pm}={1\over \omega_{\pm}}
\left[2j\pi+\cos^{-1}
\left( { a\omega^2_{\pm}+( a\gamma+b)\gamma\over c\delta(\omega_{\pm}^2+\gamma^2)}\right)\right],
\> j=0,1,2\dots
\end{equation}
and if
\begin{equation}
\sin(\omega\tau)={-\omega_{\pm}^3+(b-\gamma^2)\omega_{\pm}
\over c\delta(\omega^2_{\pm}+\gamma^2)}<0
\end{equation}
we have
\begin{equation}
\tau^j_{1,\pm}={1\over \omega_{\pm}}
\left[(2j+2)\pi-\cos^{-1}
\left( { a\omega^2_{\pm}+( a\gamma+b)\gamma\over c\delta(\omega_{\pm}^2+\gamma^2)}\right)\right],
\> j=0,1,2\dots
\end{equation}
The analogous critical time-lags  from the second factor of the characteristic equation
are given as follows. If
\begin{equation}
\sin(\omega\tau)={-\omega_{\pm}^3+(b-\gamma^2)\omega_{\pm}
\over c\delta(\omega^2_{\pm}+\gamma^2)}>0
\end{equation}
we have
\begin{equation}
\tau^j_{2,\pm}={1\over \omega_{\pm}}
\left[2j\pi+\cos^{-1}
\left( {- a\omega^2_{\pm}-( a\gamma+b)\gamma\over c\delta(\omega_{\pm}^2+\gamma^2)}\right)\right],
\> j=0,1,2\dots
\end{equation}
and if
\begin{equation}
\sin(\omega\tau)={\omega_{\pm}^3-(b-\gamma^2)\omega_{\pm}
\over c\delta(\omega^2_{\pm}+\gamma^2)}<0
\end{equation}
we have
\begin{equation}
\tau^j_{2,\pm}={1\over \omega_{\pm}}
\left[(2j+2)\pi-\cos^{-1}
\left( {- a\omega^2_{\pm}-( a\gamma+b)\gamma\over c\delta(\omega_{\pm}^2+\gamma^2)}\right)\right],
\> j=0,1,2\dots
\end{equation}

The previous formulas give  bifurcation curves in the plane
$(\tau,c)$ for fixed values of the parameters $a,b$ and $\gamma$.
We denote any bifurcation value of the time-lag by
 $\tau_c$ and add the subscripts and superscripts to specify a particular
 branch of $\tau_c(c)$.
The bifurcations are either  subcritical Hopf  on the curves
 $\tau^j_{1,-}$ and $\tau^j_{2,-}$
 leading to a disappearance of one unstable plane,
or  supercritical Hopf  on $\tau^j_{1,+}$ and $\tau^j_{2,+}$
resulting in appearance of an
 unstable plane.
The type of the bifurcation can be seen by
 calculation of the variations of the real parts ${\bf Re}\lambda$ as the time-lag
 is changed through the critical values. Again, differentiation of the characteristic equation
  gives
$$
\left({\partial \Delta_1\over \partial \lambda }\Delta_2+
\Delta_1 {\partial \Delta_2\over \partial \lambda}\right)
{d\lambda\over d\tau}
=-\left({\partial Eq_1\over \partial \tau}\Delta_2+
\Delta_1 {\partial \Delta_2\over \partial \tau}\right)
$$
and
$$
sgn\left({d{\rm Re} \lambda\over d\tau}\right)_{\tau=\tau_c}=
sgn\left\{{\rm Re} \left({d\lambda\over d\tau}\right)^{-1}\right\}_{\tau=\tau_c}=
sgn\left({2\omega^2+A\over c^2\delta^2(\omega^2+\gamma^2)}\right)
$$
Substitution of $\omega=\omega_{\pm}$ finally gives
\begin{equation}
\left({d{\rm Re} \lambda\over d\tau}\right)_{\tau=\tau_+}>0,\qquad
\left({d{\rm Re} \lambda\over d\tau}\right)_{\tau=\tau_-}<0.
\end{equation}

 Let us now discuss the zero solution of (10). Such solution would imply that:
 $c=(a\gamma+b)/\gamma\delta$. For all examples of the coupling functions that we
 have considered, like: linear, sigmoid, $\tan^{-1}$ or $\tanh$, this value of $c$
 was always larger than $c_1$, i.e. there were nonzero stable stationary points of (3),
  So we disregard such solutions
 of the characteristic equation (10), and concentrate only the Hopf bifurcations
$\lambda=\pm i\omega$.

 The bifurcation curves $\tau_c(c)$, given by (19),(21),(23) and (25), are shown in
figures 2, 3 and 4, for the first few $j=0,1,2$, and
 for the parameters $a,\gamma$ and $b$ fixed to some typical values, and
 for the coupling function with $f_0'=\delta=1$.
 Bracketed letters indicate the number of stable and unstable planes in the
 considered area
 of the $(c,\tau)$ parameter space. For example $(u^2,u)$ means two pairs of
 unstable eigenvalues of the first factor in (10) and one pair of the unstable
 eigenvalues of the second factor. Analogously, $(s,s)$ means that all
 eigenvalues have negative real parts, i.e. the stationary solution is stable.

Consider first the coupled excitable units when the coupling is
  in the range $c\in (c^{\tau},c_0)$. In this range, the condition (20) applies
   for $\omega=\omega_{+}$, and the condition (18) for $\omega=\omega_{-}$.
   Accordingly $\tau^j_{1,+}$ branches should be calculated with
     formula (21), and $\tau^j_{2,+}$ using (23). The $-$ branches
      should be computed  with (19) for $\tau^j_{1,-}$ and (25) for $\tau^j_{2,-}$.
This gives the bifurcation curves for $c\in (c^{\tau},c_0)$
presented in figure 2a. The first unstable $(c,\tau)$ domain  is
between the curves $\tau^0_{2,+}$ and $\tau^0_{2,-}$. The unstable
plane is given by $x_1=-x_2$ and $y_1=-y_2$. The corresponding
bifurcation on $\tau^0_{2,+}$ is the inverse  Hopf bifurcation
which results  in the destabilization of the stationary point and
a collapse of an unstable limit cycle. The origin of the latter is
in a global bifurcation which will be discussed in the next
section, together with the unique global attractor that exists in
this parameter domain.
 The next unstable region between $\tau^0_{1,+}$ and
 $\tau^0_{1,-}$, is bordered by a direct supercritical Hopf
 bifurcation at $\tau^0_{1,+}$ and the sub-critical Hopf
  bifurcation at  $\tau^0_{1,-}$. The unstable invariant manifold
 of the stationary point is given by $x_1=x_2,\> y_1=y_2$. The
 stable limit cycle in it supports coherent in-phase oscillations
 of the two units.
 The unstable domains bordered by the different branches start to
  overlap for sufficiently large time-lags,
 leading to multi-dimensional unstable
  manifolds of the stationary point. The global attractors for
  large time-lags are studied in the next section.

Next we consider the range of coupling $c\in(c_0,c_1)$ (see
figures 3 and 4).
  Then, for a sufficiently small $\tau>0$, there is only one pair of roots of (11)
  in the right half-plane, and all the other roots of (11) and (12)
  have negative real parts.
 There is an unstable stationary solution and the stable limit cycle.
 If $c>c_0$ but is smaller then some $c_s$,
 the Hopf bifurcation that happens for the smallest
 time lag corresponds to $\tau^0_{1,-}$. The value $c_s$ corresponds to the
 intersection of the branches  $\tau^0_{1,-}$ and $\tau^0_{2,+}$.
 If $c>c_s$ the
 first bifurcation occurs for $\tau^0_{2,+}$.
 In the parameter area below the two curves $\tau^0_{1,-}$ and $\tau^0_{2,+}$, denoted
 by $(u;s)$, the stationary solution has qualitatively the same properties
 as for $\tau=0$ i.e. it is
 unstable and has the unstable 2D manifold with
 the stable limit cycle in it. The bifurcation at $\tau^0_{1,-}$ is inverse subcritical
 Hopf, which results in a stabilization of
 the stationary solution and in a creation of an unstable limit cycle.

From the set of frames in figure 4, we see that as
$(b+\gamma)\rightarrow 0$ the
 value $c_s$ approaches $c_0$ and the $(s,s)$  domain beyond $c_0$ shrinks to nothing. In fact, in this
 singular limit, there are only positive solutions of the equation (16),
 and the stabilization of the stationary point by the time-delay can not happen.

  In order to claim that the parameter domain denoted $(s,s)$, where the stationary
 point is stable, corresponds to the phenomenon of oscillators death, the local stability
 of the stationary point is not sufficient. We need to analyze the global dynamics of (3),
 and this depends on the full form of the coupling function.

Let us briefly discuss the modifications of the presented analyzes
 that would be implied by the substitution of the coupling function of the form
 like in (3) by the diffusive or a more general coupling $f(x_1,x_2)$.
 The analysis of the linear stability in the delayed case, in particular the
 formulas for the critical time-lags and eigenvalues, would remain unchanged
 provided that the parameters $a$ and $\delta$ are changed as follows
$$ a\rightarrow  \bar a=a+\partial_1f_0\qquad \delta\rightarrow
\bar \delta=\partial_2 f_0. $$ In particular, for the diffusive
coupling, with $f(x_1,x^{\tau}_2)=(x_1-x^{\tau}_2)$,
 there will be a bounded $(s,s)$ region where the
 stationary point is stable for some finite, nonzero $\tau$ and unstable for
 $\tau=0$, and the other parameters unchanged. For example,
 if $a=0.25,b=\gamma =0.02, c=0.132$ the stationary point is stable
 for $\tau=0.85$ up to $\tau=24.5$, and  unstable for $\tau=0$ up to $\tau=0.85$
   and above $\tau=24.5$.
However, the unstable manifolds with the limit cycles for small
 $\tau$ would not be given by $x_1=x_2,\>y_1=y_2$ plane.

 Contrary to the case of coupled excitable  units, the stationary solution
  of coupled identical FitzHugh-Nagmo oscillators, with the
   same type of coupling, remains unstable
  for any value of the time-lag. Thus, there could be no oscillator death
   in the case of coupled FitzHugh-Nagmo oscillators with the
    considered type of coupling. On the other hand, it is known (\cite{Roy1}, \cite{Roy2}) that different type
    of coupling, like reverse diffusive, does lead to the amplitude death at least when the
    oscillators are near the Hopf bifurcation, i.e. $I>I_0; \>  I\approx I_0$.
  However, as we have pointed earlier, there is no Hopf bifurcation of the trivial stationary
  point of the
    excitable systems with the reverse diffusive coupling so that the stationary state
    is in this case always stable.

\section{ Global dynamics of the system with two units}

We study the global dynamics of (3) by numerical computations
 of orbits for different typical values of the parameters $(c,\tau)$ in each of the domains
 in the local bifurcation diagram of the stationary point up to moderately large values of $\tau$.
Our main interest is to determine if there is one or more then one attractors, and,
 if there are  stable oscillatory solutions, what is the dimensionality and properties
 of the oscillations.
 As examples of the coupling we used different functions,
 with the same qualitative conclusions, and for illustration we use
$f(x)=\tan^{-1}(x)$. On the other hand, diffusive coupling
$f(x_1,x^{\tau}_2)=(x_1-x^{\tau}_2)$ implies
 quite different global dynamics, which we shall indicate at the end of the section.

Before presenting the results let us
 comment on the initial conditions for the system (3) that we used in calculations.
In order to uniquely fix a solution of the delay-differential equations (3) one needs
 to specify the solution on the interval $[-\tau,0]$. In our calculations we always use
 a physically plausible initial functions on $[-\tau,0]$ given by solutions of the
 equations (3) with $c=0$. Thus, before the period $\tau$ has
elapsed each of the units was evolving independently of the other unit.

Firstly we discuss the global dynamics for the coupling constant
$c<c^{\tau}$, i.e. when the trivial stationary solution of the
 whole system is
stable for any time-lag. Intuitively, one would expect that if the
time-lag $\tau$ is such that $2\tau$ is smaller than the minimal
time needed for an excitable orbit to approach the stable
stationary solution, then the coupling would just induce both
units to fire one spike each, with some time-delay, and relax to
 the stationary solution. However, if $2\tau$ is larger than the
  indicated minimal time, then
 the excitation of one of the units would arrive just in time to
  kick the state of the
 other unit from close to the stationary into the excitable
 domain, even if the coupling constant is rather small. Thus,  the
 excitable orbit of the coupled system becomes periodic.
 Nevertheless the equilibrium state remains a stable stationary
 solution.  The system as a whole is bi-stable excitable with periodically spiking
 excitations. This picture is confirmed by numerical computations.
 Two typical orbits are illustrated in figure 2b.
 The periodic orbits are supported on the  stable limit cycle.
 The latter is created in a global fold limit cycle bifurcation, together with
  an unstable one. As expected, the motion on the limit cycle is
  coherent and out-of-phase, with the frequency that increases with the
   time-lag.

 Qualitatively different global dynamical pictures can occur for
the coupling constant in the range $(c^{\tau},c_0)$ and for
various $\tau>0$. In the domain of the $(c,\tau)$ plane bordered
by  $\tau^0_{2,+}$ and $\tau^0_{2,-}$, there is only one global
attractor given by the large stable limit cycle, since
 the value $\tau^0_{2,+}$ is above the critical time-lag of the
 global fold limit cycle bifurcation discussed in the previous
 paragraph, and the stationary point is unstable (see figure 2c). The large cycle is
not affected qualitatively by the local bifurcation on
$\tau^0_{2,+}$
 or $\tau^0_{2,-}$, so the dynamics on it is given by coherent out
 of phase oscillations of the coupled units.
   If the domain is
 entered by crossing $\tau^0_{2,+}$ the
 unstable limit cycle collapses on the stationary solution, and if
 the domain is left through $\tau^0_{2,-}$ yet another unstable
 limit cycle is created.

 The next unstable domain, with $c$ still in $(c^{\tau},c_0)$ is
 bounded by $\tau^0_{1,+}$ from below and $\tau^0_{1,-}$ from above.
 The supercritical Hopf bifurcation on $\tau^0_{1,+}$ results in a
 creation of a 2D unstable manifold of the stationary point given by $x_1=x_2$ and
 $y_1=y_2$. In it, there is a stable limit cycle supporting
 in-phase coherent oscillations. However, the large limit cycle with
  out-of-phase oscillations is not affected by the local bifurcation at
  $\tau^0_{1,+}$, so that is this domain the dynamics is bi-stable
  with two limit cycle attractors. This is illustrated in
  figure 2d.

 Next we consider possible attractors for the coupling constant beyond $c_0$, i.e.
  when instantaneously coupled units beheve as two limit cycle oscillators.  Qualitatively
different dynamics corresponding to different domains in
$(c,\tau)$, are illustrated  in figure 5. Frame 5b corresponds
 to a typical values of $(c,\tau)$ in the domain $(u,s)$. We have not found any global
 bifurcation that would occur as $(c,\tau)$ are varied inside the $(u,s)$ region.
 The dynamics is characterized by the 2D unstable manifold of the stationary
 solution, given by
$x_1=x_2,\> y_1=y_2$ (see fig 5a). There is a globally stable limit cycle inside this manifold.
 Oscillations of $x_1$ and $x_2$ on this limit cycle are obviously in phase.

Two frames, 5c and 5d, correspond to the situations in $(s,s)$
 with one stable stationary solution but with two  globally different dynamics.
 In 5c the system is bi-stable. There is the stable stationary solution and the
 stable large limit cycle in the plane $x_1=x_2,\> y_1=y_2$. There is also a small
 unstable limit cycle, that is created in the inverse sub-critical Hopf bifurcation
 at $\tau=\tau^0_{1,-}$. This cycle acts as a threshold between the sub-excited damped
 oscillations and periodic synchronous spiking of both units. As $\tau$ is increased,
 but still for $(c,\tau)\in (s,s)$, the unstable limit cycle approaches the stable one,
 and they disappear in a fold limit cycle bifurcation, which occurs in the invariant
 plane  $x_1=x_2,\> y_1=y_2$. Thus, there is a parameter region inside $(s,s)$ where
 $(0,0,0,0)$ is globally stable, and
that corresponds to the death of the identical oscillators.
 However, let us stress again that the global dynamics for the
parameters in the domain $(s,s)$ could correspond to either
 spiking excitability, with
 sub-threshold dumped oscillations and sup-threshold periodic spiking, or to the
 death of oscillators. In the latter regime the whole  system
 is  excitable with the stationary point
 as the only attractor.

The global dynamics above the curve $\tau^0_{2,+}$ is
characterized by one large limit cycle as the global attractor.
The oscillations on it are coherent and out-of-phase.
 The same type of the global attractor occur above the critical line
 $\tau^{1}_{1,+}$ as illustrated in frames 5e and 5f.
 The oscillations are further illustrated
 in figure 6b by plotting the limit cycle as seen in the coordinates $(x_1-x_2,y_1-y_2)$.
 Convergence to the limit cycle is much slower than in the case of the symmetric oscillators
 that occurs for smaller $\tau$, illustrated in figure 6a.
In fact, in all domains up to a quite large value of the time-lag
the global attractor is a stable limit cycle (could be imbedded in
a multi-dimensional
 unstable manifold
 for larger $\tau$) which  supports
 asymmetric phase shifted oscillations of $x_1$ and $x_2$.
 However, there are domain for larger values of the time-lag, for example for
 $\tau=55$ and any $c\in(c_0,c_1)$, where
 the global attractor is the symmetric limit cycle, with coherent
 and in-phase oscillations.

 It should be pointed out that, for all larger time-lags up to quite
 large values, equal to several refractory times of the non-coupled
 units, the attractor is always a limit cycle.
On the limit cycle, all variables
 oscillate with the same frequency, and could be either symmetric
 or phase shifted. The two regimes interchange as the time-lag is increased.
  These are the only two possible stable attracting patterns,
 despite the large dimensionality of the unstable manifold of the
 stationary point. It should be pointed out that in the case of
 identical FitzHugh-Nagumo oscillators with the same coupling various types
  of quasi-periodic attractors occur for moderate values of the
  time-lag. However, also in this case, stronger coupling and
  larger time-lags imply synchronization, either identical or phase
  shifted, like for the coupled excitable systems.
The dynamics  for time-lags much larger than the refractory time
has not been  systematically studied.

We now briefly comment on  the dynamics in the case of the diffusive coupling. As stated before,
 at some $c$ and for $\tau$ zero or small, the only attractor is the stable limit cycle,
 with coherent and phase-shifted oscillations of the two units. The time delay can stabilize
 the trivial stationary point, but the system remains bi-stable with the limit cycle and the
 stationary point as the attractors,
  for all values of $(c,\tau)$ in $(s,s)$ domain. The phenomenon of oscillator death does
 not occur in the case of the diffusive coupling.

\section{ $N>2$ lattice}

The goal of this section is to present numerical evidence that for
some common types of lattices with $N>2$ there are  regions in the
parameter plane $(c,\tau)$ analogous to $(u,s),(s,s),(u,u)\dots$
 in figures 2 and 3.
We have analyzed examples of systems of identical FitzHugh-Nagumo
excitable units arranged in linear or circular lattices, with
unidirectional
 or bi-directional symmetrical coupling by few typical coupling functions. Lattices of the size
 $N=10,20$ and $N=30$ have been  studied systematically.

The conclusions are illustrated using the following model:
\begin{eqnarray}
\dot x_i & = & -x_i^3 + (a + 1)x_i^2 - ax_i -y_i+cf(x_{i-1}^{\tau})+cf(x_{i+1}^{\tau}),\nonumber \\
\dot y_i & = & bx_i - \gamma y_i,\qquad i=2,\dots N-1,
\end{eqnarray}
\begin{eqnarray}
\dot x_{1,N} & = & -x_{1,N}^3 + (a + 1)x_{1,N}^2 - ax_{1,N} -y_{1,N}+cf(x_{2,N-1}^{\tau}),\nonumber \\
\dot y_{1,N} & = & bx_{1,N} - \gamma y_{1,N},
\end{eqnarray}
where the coupling is given by $f(x)=\tan^{-1}(x)$, and $N=20$.

Firstly, in the case of instantaneously coupled units, there is
the Hopf bifurcation at some $c=c_0$.
 As in $N=2$, for the coupling constant below some
$c^{\tau}<c_0$ and any time-lag the trivial stationary point is
stable. If the time-lag is sufficiently large, there is also the
stable large limit cycle. On it, all units oscillate coherently.
However, the nearest neighbours oscillate exactly anti-phase, so
that two clusters are formed.

The coupling above the threshold $c_0$, and for small
 time lags, leads to the appearance of a globally stable limit cycle representing
 synchronous oscillations in the plane given by
 $x_2=\dots =x_{N-1},\>y_2=\dots =y_{N-1}$ and $x_1=x_{N},\>y_2=y_{N}$,
 ( $(u,s)$ region). As expected, the synchronization period could be quite large if the
 value of the coupling constant is near $c_0$, i.e. when each of the units
 is near the Hopf bifurcation.

Increasing the time lag leads to the inverse Hopf bifurcation.
 For any $c$ in some interval $(c_0,c_s)$ we have been able to find intervals
 of time-lags $(\tau_c^{-},\tau_c^{+})$ that
 correspond to bi-stability or to death of all $N$ oscillators ($(s,s)$ region),
 illustrated in figure 7. The same figures illustrate the
 attractors in the dynamics of any of the identical neurons.
 Again, the inverse Hopf  and the subsequent fold limit
 cycle bifurcations, due to increasing time-delay,  are responsible for the amplitude death in
 the systems of the form (27). On the contrary, the stationary
 point of a lattice like (27) with the same coupling
  but with Fichugh-Nagumo oscillators is always unstable.

  Larger time lags do not change the topological nature of the
  attractor. It is always a limit cycle, but the synchronization
 pattern between the coherent oscillations of the units depends
 on $\tau$.
 Non-symmetric oscillations with equal frequencies appear after long transients, as is
  illustrated in figure 8a. Dynamics in the transient period can be quite
 complicated. Properties of the asymptotic synchronization patterns could
 depend on the geometry of the lattice.

Existence of the presented types of dynamics, and the order of their appearance as
 $(c,\tau)$ are varied, was confirmed in all examples that we have studied. We
 conjecture
 that qualitative properties of the dynamics of all small 1D lattices with nearest-neighbor
 delayed coupling of the form like in the equations (27) between the
 identical FitzHugh-Nagumo excitable
 systems are the same, at least for not very large values of the time lag,
 in the sense that the
 same types of bifurcations appear and determine the dynamics.

\section{ Summary and discussion}

We have studied small lattices of excitable identical units with time-delayed coupling,
 where each unit is given by the excitable FitzHugh-Nagumo model. The coupling is
 always between the voltages of the nearest neighbors, but could be of a quite general
 form. Our primary interest was in the bifurcations and the typical dynamics that occur
 for time-lags which are not very large on the
 time-scale set up by the refractory or the inter-spike
 period. Detailed study,  in the case of only two units, of the local stability and bifurcations of
 the stationary solution suggested, but does not uniquely  determine, the possible global
 bifurcations and dynamics. These are studied numerically.

There are only few possible types of dynamics, at least for time
lags as large as several refractory times. For small coupling
constants
 and small time-lags there is only one attractor in the form of
 the stable stationary solution. The whole system beehives qualitatively
  as the simple excitable.
  Relatively small coupling
 constants $c<c^{\tau}$ and sufficiently large time-lag result in
 the limit cycle attractor co-existent    with the stable
 stationary solution. The whole system is bi-stable with spiking
 excitability. The oscillations on the limit cycle are coherent
 and out-of-phase. For the coupling constant above $c^{\tau}$
  the sequence of Hopf bifurcations due to the time-delay of the stationary
 solution are possible. For $c\in(c^{\tau},c_0)$ and small time
 lags the stable stationary point is the only attractor, but as
 time-lag is increased the system could be either bi-stable or
 there could be only one attractor in the form of the limit cycle.
  The bi-stability is manifested either in the form of the
 stable stationary solution and the stable limit cycle, or could
 be in the form of two stable limit cycles (one in-phase and one
 out-of-phase). The interval  $c\in(c^{\tau},c_0)$ is rather a small
 part of the $c$ values for which there is only one stationary
 solution.
  For $c\approx c_0$ each of the instantaneously coupled units is
 near a direct super-critical Hopf bifurcation, but as soon as
 $c-c_0>0$ is bigger then some quite small $\epsilon_0$
 the resulting limit cycle has quite large radius and the harmonics become influential,
 unlike in the case of the Hopf limit cycle.  Increasing the time-lag $\tau$ could lead to
 stabilization of the stationary point, via indirect sub-critical Hopf, resulting
 in a bi-stable dynamics, with a stable stationary point, small unstable limit cycle
 as a threshold, and a large stable limit cycle. Further increasing $\tau$ leads to
 a fold limit cycle bifurcation, in which the unstable and the stable limit  cycles disappear,
 and the stationary point remains the only attractor. In this, oscillator death regime,
 the system again displays the simplest form of excitability,
 like in the case of the weak coupling $c<c_0$ and zero or small
 time lags.
  Still further increase of the time-lag $\tau$ leads to the
 super-critical Hopf. The  oscillations on the limit cycle are coherent but
 are phase shifted, and the oscillators need not have the same amplitude.
 Described sequence of bifurcations happens for time lags that are all small,
 up to $10 \%$ , with respect
 to the refractory period of the single isolated unit. Further
 increase of the time-lag leads to more dimensional unstable
 manifold of the stationary solution. However, the
  global attractor is always a simple limit cycle. The asymptotic
  dynamics is always coherent, and is either in-phase or phase
  shifted.
  Unlike the case of coupled
  FitzHugh-Nagumo oscillators, nothing more complicated than the
  limit cycle could be the attractor of the coupled excitable
    FitzHugh-Nagumo systems.

 Our analyzes shows that the most common type of
 excitations of the whole system, in response to an impulse submitted to either of
  the units, is in the form of coherent
 out-of-phase oscillations. However, if the transmission is sufficiently strong
  and for moderately large transmission delays of signals between the
  units, the compound system would respond by synchronous in-phase
 oscillations. Furthermore, our results
 suggest  that relatively small but non-zero time-delay together
 with sufficiently strong interaction could result in a simple
 excitable behaviour of the compound system. For such values of
 the parameters the system would operate as a powerful amplifier
 of a quite small impulse administered to its single unit.
 Due to the particular model of the excitable system and to the type
of coupling that we have studied in detail, the most relevant
possible application of our results is in modeling
 coupled neurons. In fact, relatively recent experiments and
 analyzes \cite {Clay} show that the FitzHugh-Nagumo equations,
 despite the common opinion, might represent better qualitative
 model of an excitable neuron than the more detailed Hodgkin-Huxly
 system. Our results indicate that the fine tuning between the
 synaptic coupling and delay could lead to the in-phase
 synchronous operation of a collection of neurons.

 Although there is a quite substantial amount of work done on the systems of  dynamical
 units with the delayed coupling, such systems are comparatively much less studied than
 the corresponding systems with the instantaneous coupling. For the purpose of comparison
 with our work, we shall try to classify the existing contributions into typical groups.

Firstly, we consider  the model and the results presented in \cite{Roy1},\cite{Roy2}.
 In these papers, a network of $N=2$ and $N>2$ oscillators described by the equations of
 the normal form of the Hopf bifurcations with delayed inverse diffusive coupling is studied.
 At zero coupling, and/or for small time lags, all oscillators have small limit cycles
 just created by the direct Hopf bifurcation.
 It is shown that the time-delay can lead to the stabilization of the
 trivial  stationary point, even for the
 identical oscillators, which is interpreted as the amplitude death.
 Analogously, in our case, the
 Hopf direct bifurcation is responsible for the appearance of the oscillations when the
 excitable units are instantaneously coupled, and the time-delay leads to stabilization of the stationary
 point. The death of oscillations in our case appears after the fold bifurcation of the
 stable and the unstable limit cycles, which are created in the
 same plane. The oscillator death occurs only in a domain in the $(c,\tau)$
 parameter space smaller that the domain of the stability of the stationary point.

Next, we compare our model and the results
 with those that appear in the studies of the delayed coupled relaxation
 oscillators, for example in : \cite{Fox},\cite{relax2}.
 In these studies, each unit
 is a relaxation oscillator, and the primary objective of the analysis are the periodic
 orbits that appear in the delayed coupled system. Singular approximation,
 or an approximate or numerical construction of the Poincare map,
 are used to analyze various types of synchronous or asynchronous oscillations.
The phenomenon of the oscillator death was not observed (  \cite{Ermentrout1}).
In our case, the noniteracting units are not oscillators and the oscillations are
 introduced by coupling, via the Hopf bifurcation.
 The domain of parameters $(c,\tau)$ that implies oscillator death shrinks to nothing
 in the singular limit $(b+\gamma)\rightarrow 0$. Furthermore, the FitzHugh-Nagumo
 model is type II excitable, which reflects in the type of bifurcations that
 might occur in the coupled systems.

Less directly related to our work is the analyzes of the influence
of the time delay in the
 systems of coupled phase oscillators.
In fact, if the rate of attraction to the limit cycles
 of two voltage-coupled neural oscillators is sufficiently strong the dynamics
can be described by the coupled phase oscillators.  The coupling
 between the phases mimics the voltage coupling, and is not of the diffusive type.
 The phenomenon of oscillator death in such instantaneously coupled phase oscillators
 was studied for example in \cite{Ermentrout1}. The influence of time-delay in
 coupled phase oscillators was studied for example in
  \cite{phase1} and \cite{phase2} (and also in \cite {Ermentrout1}), where it is shown that
 the time-delay can not introduce significant changes into the dynamics of a class
 of such systems \cite{phase1}, unless the time-lag is of the order of several oscillation
 periods \cite{phase2}.
Independently of neuronal models, collective
 behavior of the phase coupled (phase) oscillators
 with time-delayed coupling, have been studied using the dynamical
\cite{Schuster1},\cite{Schuster2},\cite{Nakamura},\cite{Kim}), or statistical \cite{Strogatz}
 methods.
 In our case, the coupling is between the voltages, could be of a
 quite general form, and
all analyzes and the observed phenomena occur already for quite
small time lags.

 Finally, the influence of time delay has been studied in the Cohen-Grasberger-Hopfield (CGH)
 type of neural networks, as early as in 1967 \cite{Grossberg}. More recent references are
 for example
\cite{Leung}, \cite{Wang}, and for small networks \cite{Campbel3},\cite{Campbel4},
 \cite{Campbel2},
 (see also \cite{Campbel1} and the references there in).
 In the non-delayed case, the stability of the stationary point in such
 networks is proven using an energy-Lyapunov function.
 Using the corresponding Lyapunov functional in the delayed case, it was shown in \cite{Wang},
 that the stationary point remains globally stable for sufficiently small time lags.
 On the other hand, destabilization of the stationary point occurs via the Hopf bifurcation,
 as was shown in \cite{Campbel3},\cite{Campbel4},
 for the networks with $N=2$ and $N=3$ units, and
 multiple time-delayed coupling.

 As is seen, the model treated here, and our results,
 have some features common with few other models.
 As in the CGH networks, each isolated unit has a globally stable stationary point, and
 the time-delayed weak coupling does nothing to the dynamics, provided that the time-lag is
  sufficiently small.
  If the coupling is
 strong enough, the system behaves either as a collection of near-Hopf oscillators, or as
 a collection of relaxation oscillators. Death of oscillators due to time-delay
is observed in both types of dynamics, although the phenomenon happens for a smaller
 range of time-lags if the system behaves as a collection of relaxation oscillators.

 Let us finally mention few related questions that we shall study in the future.
Firstly, it should be interesting to see if the systems of slightly different units share
 the same type of dynamics. Secondly, examples of type I excitable (and not oscillatory)
 systems coupled with time-delay should be analyzed, in order to underline the
 role of the type of excitability. Finally,
 the external pulse perturbations, like  for example in
\cite{Roy3}, \cite{Kitajima},\cite{Combes}, could introduce different transitions
 from excitability to the oscillatory regime, and the consequent changes in the dynamics
 due to time-delay should be analyzed.
  \vskip 1cm
{\bf Acknowledgements}
This work is supported by the Serbian Ministry of Science contract No. 1443.
\vskip 2cm

\section{ Appendix}

We start with the characteristic function (10) in the form:
 $$
 \phi(z)=\left[ (z+\gamma)(z+a)+b\right]^2 -c^2\delta^2(z+\gamma)^2
 \exp(-2z\tau),
$$
 and consider the following expression:
$$
 f(z)={\phi(z)\over P_4(z)}=1-{c^2\delta^2(z+\gamma)^2\over
P_4(z)} \exp(-2z\tau), $$
 where
$P_4(z)=\left[(z+\gamma)(z+c)+b\right]^2$ is actually the
characteristic function of the single non-coupled unit.

Consider the contour $C_R$ in the complex half-plane
 ${\bf Re}z>0$, formed by the segment $[-iR,iR]$ of the imaginary
 axis and the semi-circle with the radius $R$ centered at the
 origin. As the condition $4(b/\gamma)>(a-1)^2$, equivalent to
  the existence of a unique stationary solution,
  is by assumption always satisfied, the polynomial $P_4(z)$ has
  no zeros in the half-plane ${\bf Re}z>0$. In that case, the
  number of poles of $f(z)$ is $P_c=0$. Using the argument
  principle we infer the number of zeros $N_{C_R}$ of $f(z)$.
  If $\lim_{R\rightarrow \infty} N_{C_R}=0$ then all the roots of
  the characteristic function $\phi(z)$ satisfy ${\bf Re}z>0$.
 Thus, we need to find the conditions on the
  parameters $a,b,\gamma$ and $c$, such that the image of the contour $C_R$ when
 $R\rightarrow\infty$ by the function $f(z)$ does not encircle the point
 $z=0$. Then the variation of the argument is zero, so that $\lim_{R\rightarrow \infty}
 N_{C_R}=0$, and consequently the zeros of the characteristic
 function satisfy ${\bf Re}z>0$ for any $\tau$.
This is the essence of the amplitude-phase method ( see for
  example \cite{Norkin}.)

It is enough to consider the image of the segment $[-iR,iR]$ by
the function
 $$
\omega_{\tau}(z)\equiv {c^2\delta^2(z+\gamma)^2\over P_4(z)}
 \exp(-2z\tau),
 $$
 or, in fact, by just $\omega_{0}(z)$ since $|\omega_{0}(iy)|< 1$
 if and only if $|\omega_{\tau}(iy)|< 1$, and the image of the
 semi-circle shrinks to a point as $R\rightarrow \infty$.

 Since
 $$
 |\omega_{0}(iy)|=|{ c\delta(iy+\gamma)\over
 (iy+\gamma)(iy+a)+b}|^2={c^2\delta^2(\gamma^2+y^2)\over
 y^4+(a^2+\gamma^2-2b)y^2+(a\gamma+b)^2}
 $$
 we obtain that $|\omega_{0}(z)|< 1$ is equivalent with
 $$
 y^4+Ay^2+B>0
 $$
 where $A$ and $B$ are given by the same formula as in (15), i.e.
$$
A= a^2+\gamma^2-2b-c^2\delta^2,\quad {\rm and} \quad
B=(a\gamma+b)^2-c^2\delta^2\gamma^2.
$$

For the parameters such that $b>\gamma^2$ and
$4(b/\gamma)>(a-1)^2$ the above condition is equivalent to
 $$
 c<\sqrt{ {a^2\gamma^2-2b+2
  \sqrt{b(2\gamma^2+2a\gamma+b^2)}\over \delta^2}}.
$$
 The right side is the critical value that we denoted $c^{\tau}$ in the main
 text.

\newpage
{\bf FIGURE CAPTIONS}

1){\bf Figure 1} The figure illustrates continuous transition of the limit cycles
 from near Hopf to that of a relaxation oscillator for the coupled system with $\tau=0$.
 The fixed parameters are: $a=0.25, b=\gamma=0.02$.
 The cycle is created at $c=0.27$ and the smallest cycle on the figure is for
 $c=0.2702$, the next to the largest for $c=0.27048$ and the largest for $0.3$.

2) {\bf Figure 2} The figures illustrate typical dynamics below
$c_0=0.27$ for $a=0.25, b=\gamma=0.02$: a)  First few branches of
the bifurcation curves $\tau_c(c)$ given by equations
(19),(21),(23) and (24),
 for the parameters $a=0.25,\>b=0.02,\> \gamma=0.02$ and $c<c_0$;
 b) Examples of quickly  relaxing (1) and periodic excitable (2)
 orbits (projections on $(x_1,y_1)$; (c) Projection of the global attractor
  limit cycle on $(x_1,x_2)$ in the domain $(s,u)$;
  (d) Projection of the two
  limit cycle attractors on $(x_1,x_2)$ in the domain $(u,u)$.

 3) {\bf Figure 3}: First few branches of the bifurcation curves
$\tau_c(c)$ given by equations (19),(21),(23) and (24),
 for the parameters $a=0.25,\>b=0.02,\> \gamma=0.02$ and $c>c_0$.

4) {\bf Figure 4}: The same as figure 3, but for few fixed values
of $a,b,\gamma$:
 a)$a=0.25,\>b=0.005,\> \gamma=0.005$; b)$a=0.25,\>b=0.003,\> \gamma=0.003$;
 c)$a=0.25,\>b=0.0015,\> \gamma=0.0015$; d)$a=0.25,\>b=0.00075,\> \gamma=0.00075$

5){\bf Figure 5}:  Phase portraits in  $(x_1,y_1)$ (a,b,c,d,e)
plane, or $(x_1,x_2)$ plane (f). The initial points if there are
different orbits are
 indicated by numbers.
  The fixed parameters are $a=0.25,\>b=0.02,\> \gamma=0.02, c=0.3$,
 except in (a) where $c=0$.   The time-lag is: (a),(b) $\tau=0$; (c) $\tau=4$;
(d)$\tau=6$; (e),(f) $\tau=27$. Dynamics illustrated in (e) and (f) is typical
 also for other values of $(c,\tau)$ above $\tau^1_{1,+}$ curve.

 6) {\bf Figure 6}: (a) The asymptotic state is symmetric in any domain
 below $\tau^1_{1,+}$
curve in figure 3. In domains $(u,s)$, $(s,u)$ or $(u,u)$ the
symmetric state are
 the synchronous oscillations, and in $(s,s)$ the stable stationary point.
  (b) The asymptotic  synchronous
 oscillations are not symmetric for $(c,\tau)$ above the curve $\tau^1_{1,+}$ in figure 3,
 as is illustrated for a pair $(c,\tau) \in (u,u)$, but become symmetric for $\tau\geq 55$
  (not illustrated, see the main text).

7)  {\bf Figure 7}: Bi-stability (a) and oscillator death in the
lattice (27) with $a=0.25,b=0.02,\gamma=0.02$ and (a)
$(c,\tau)=(0.16,4)$ or (b) $(c,\tau)=(0.16,6)$.

8) {\bf Figure 8}: Asymptotic states of the lattice (27) for
 $(c,\tau)=(0.16,15)$ are
 coherent oscillations but with a fixed time lag, represented by the projection of the limit
 cycle on the $(x_4-x_{15},y_4-y_{15})$ plane in the frame (b). For such a small $c$
 the synchronization period is more then 10 times larger than the characteristic
 period, as is illustrated in frame (b) with the time dependence of the time-series
 $x_4(t)$ and $x_{15}(t)$

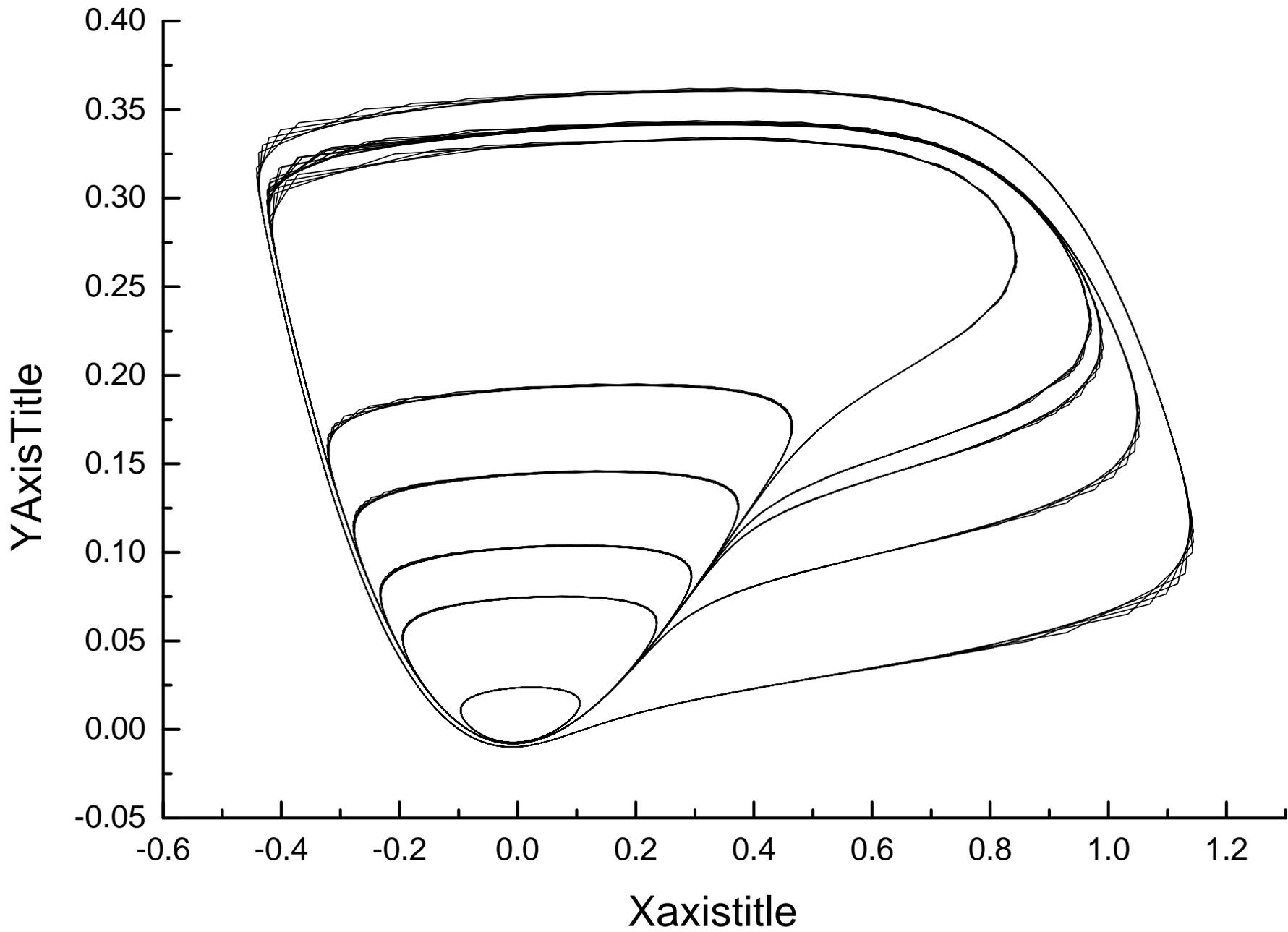
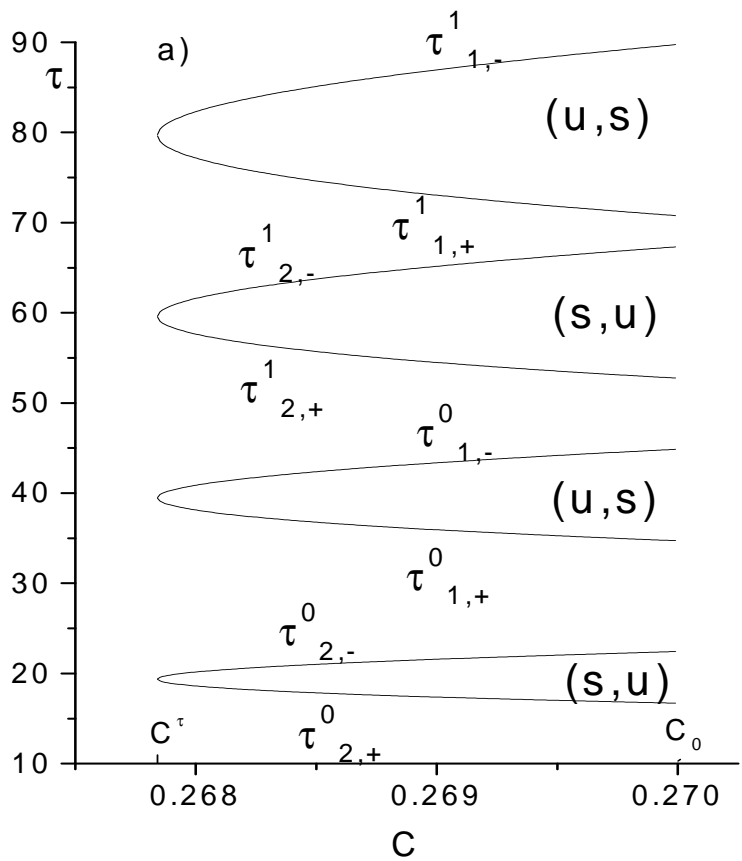
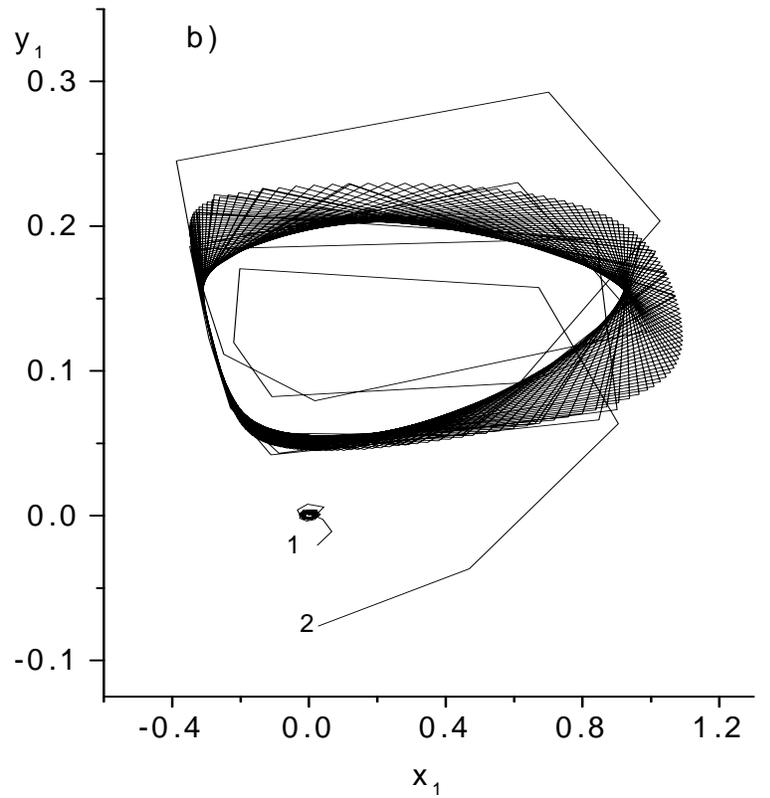
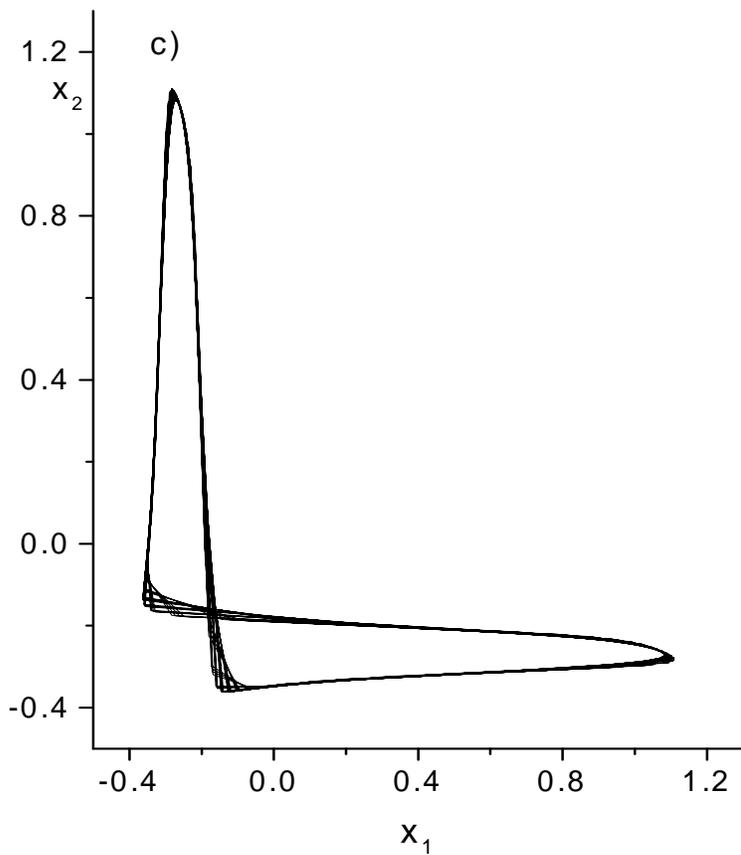
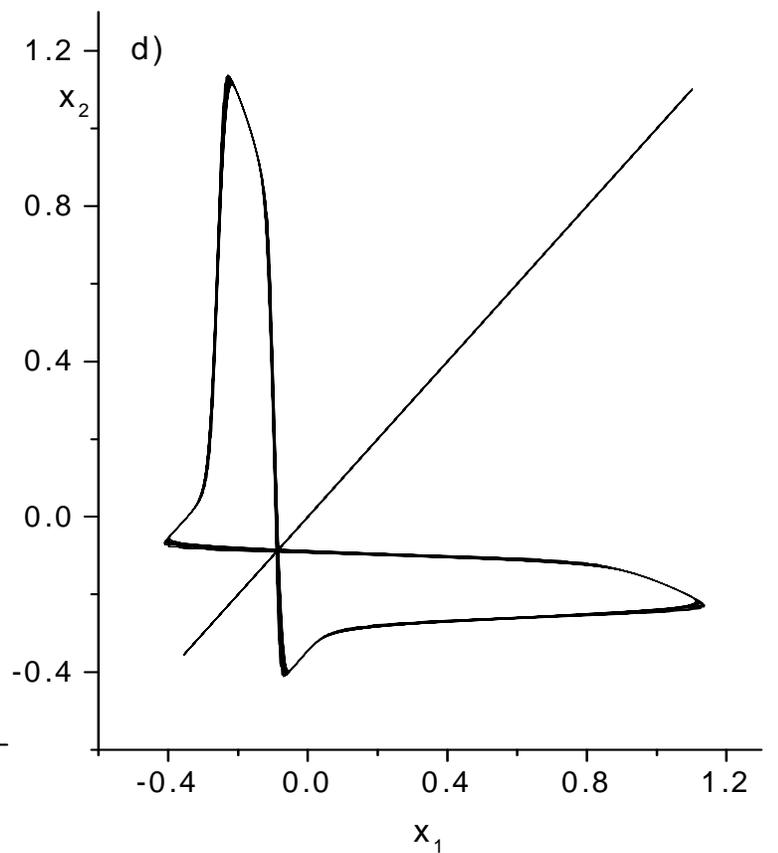
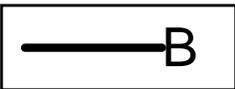
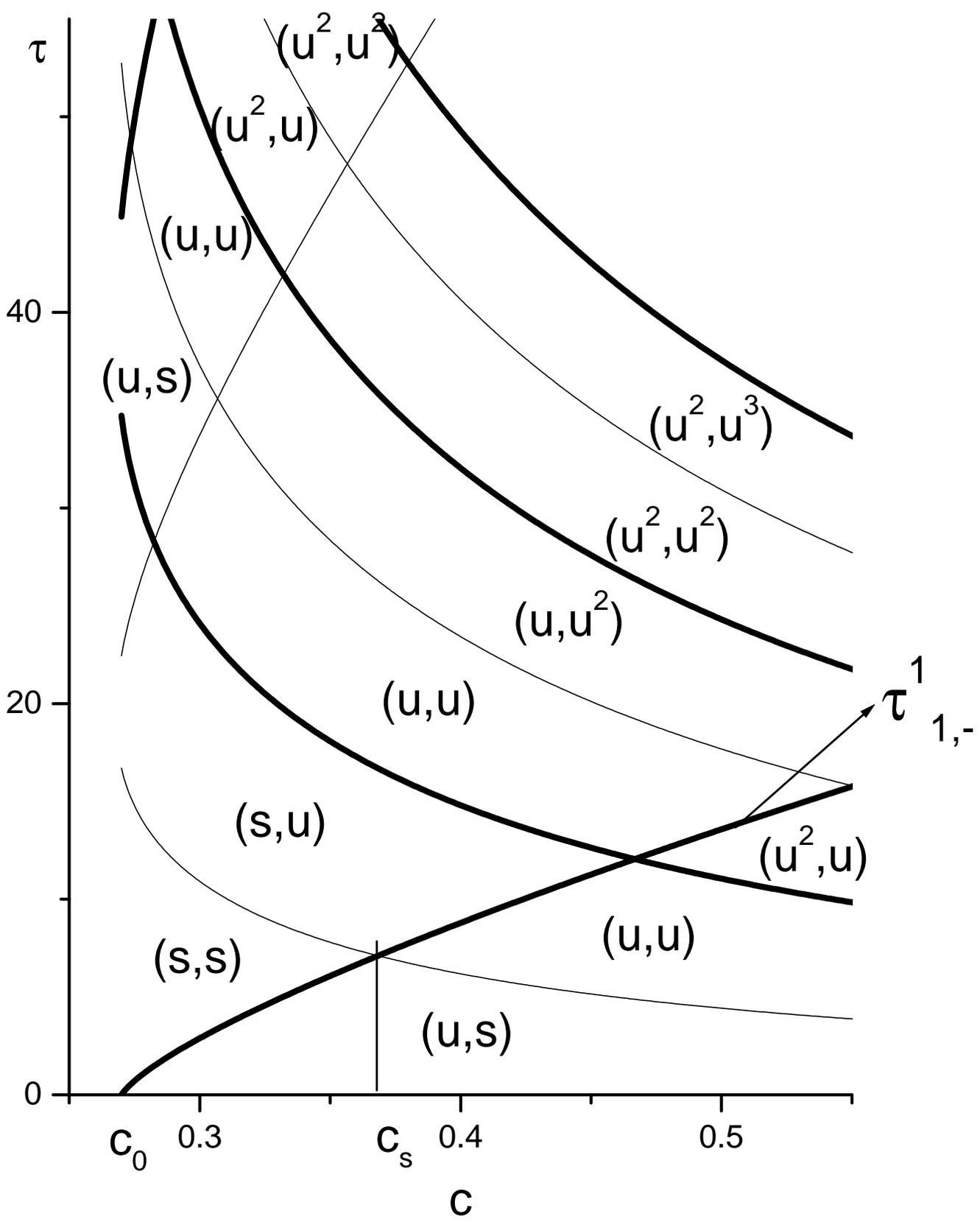
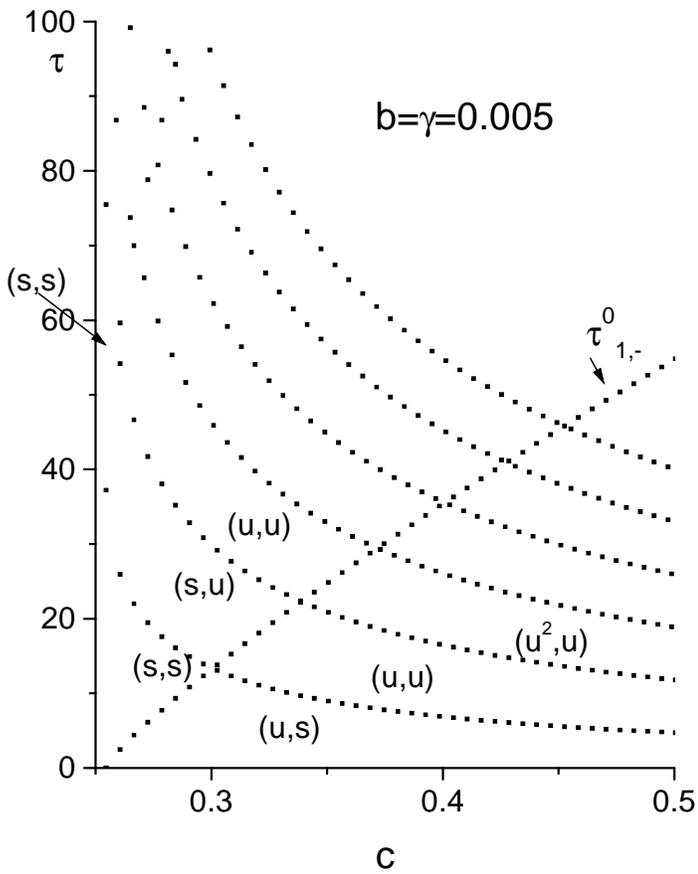
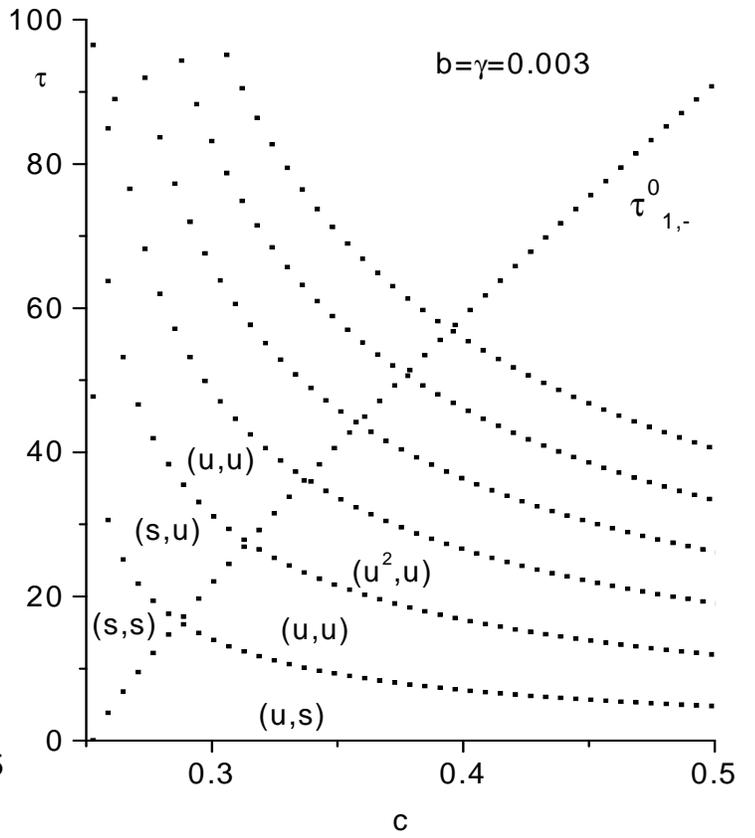
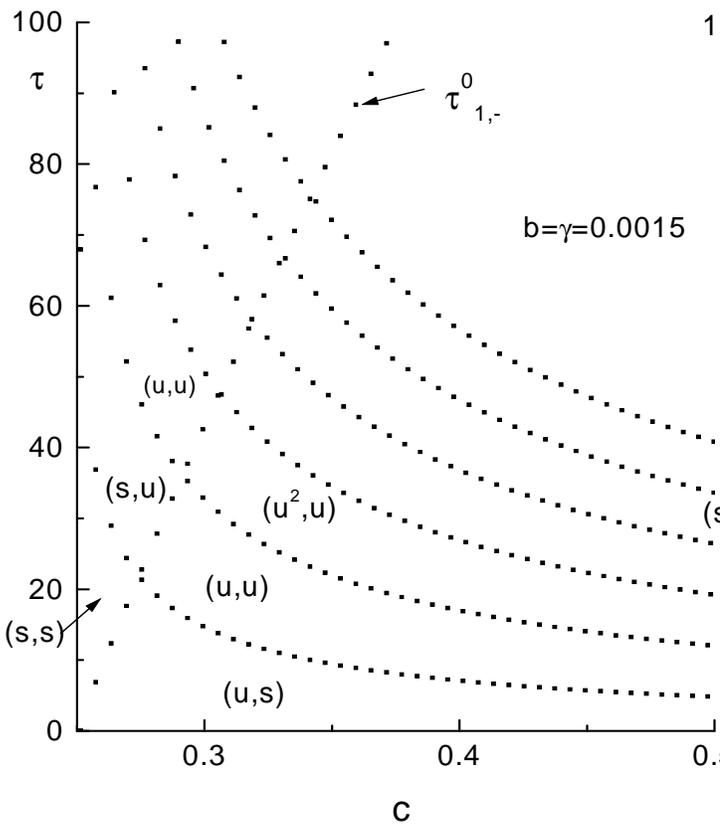
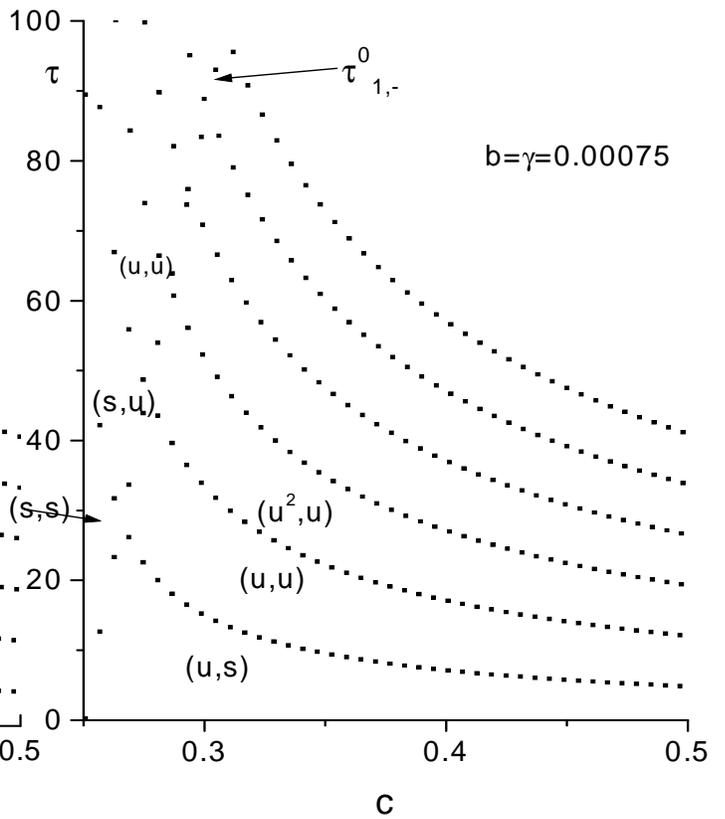
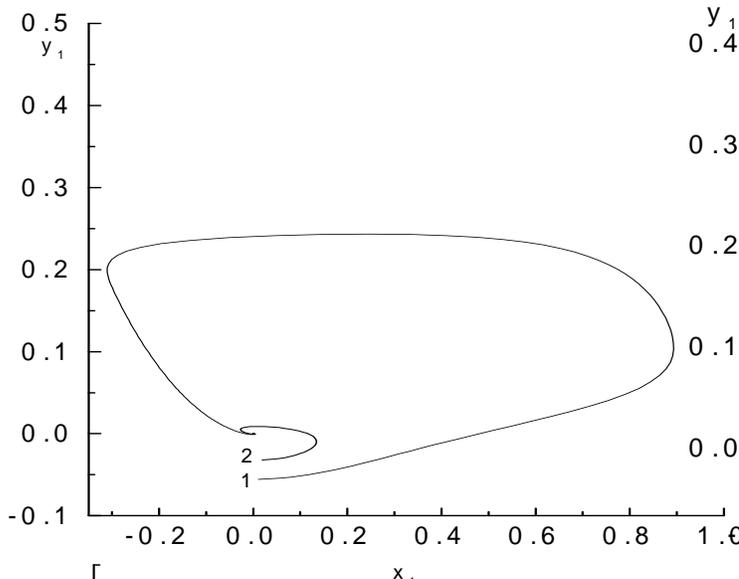
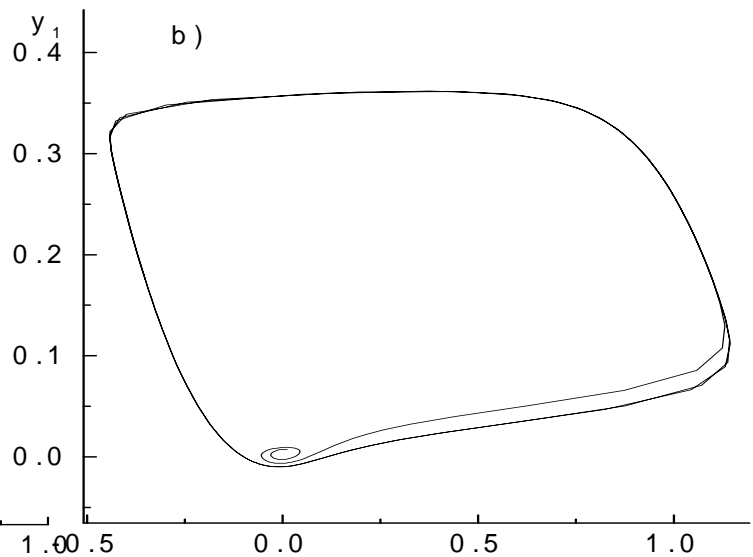
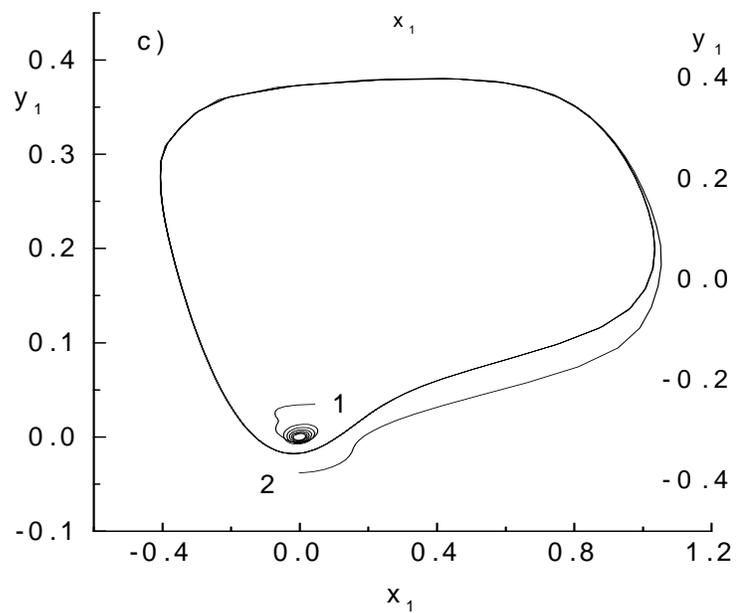
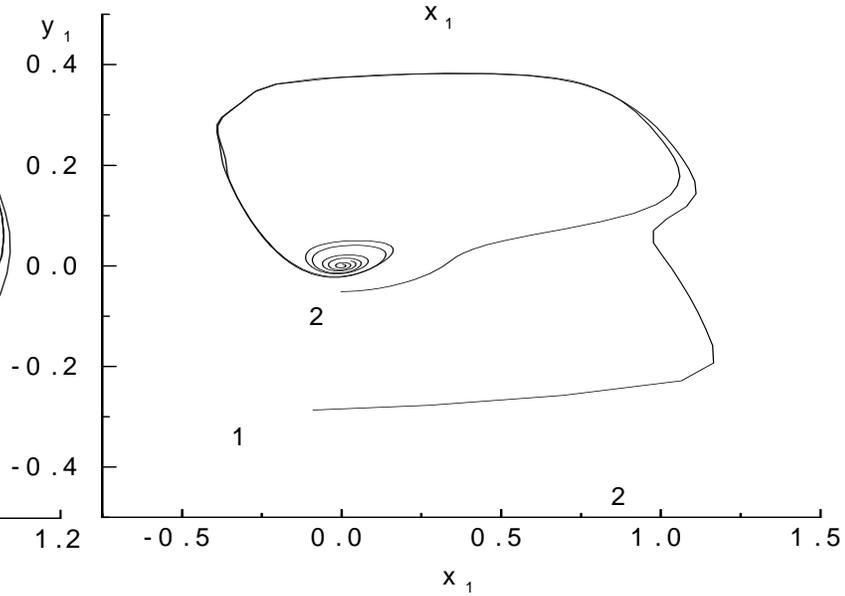
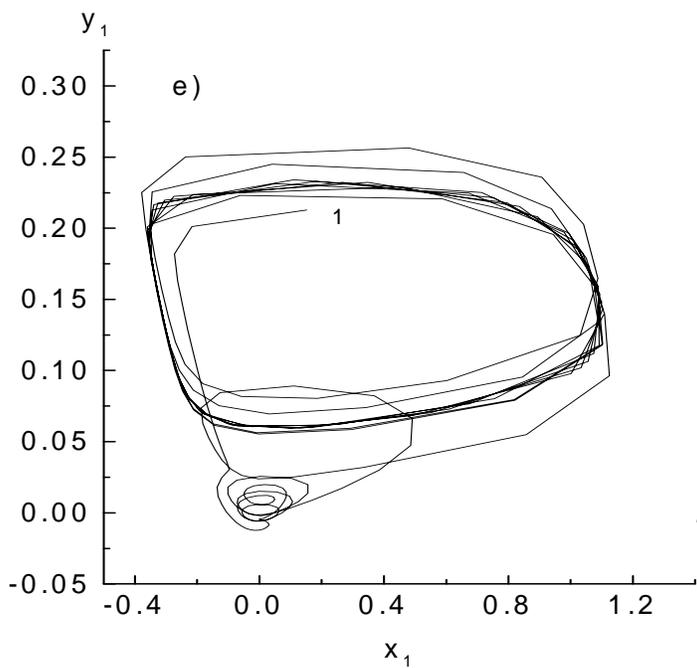
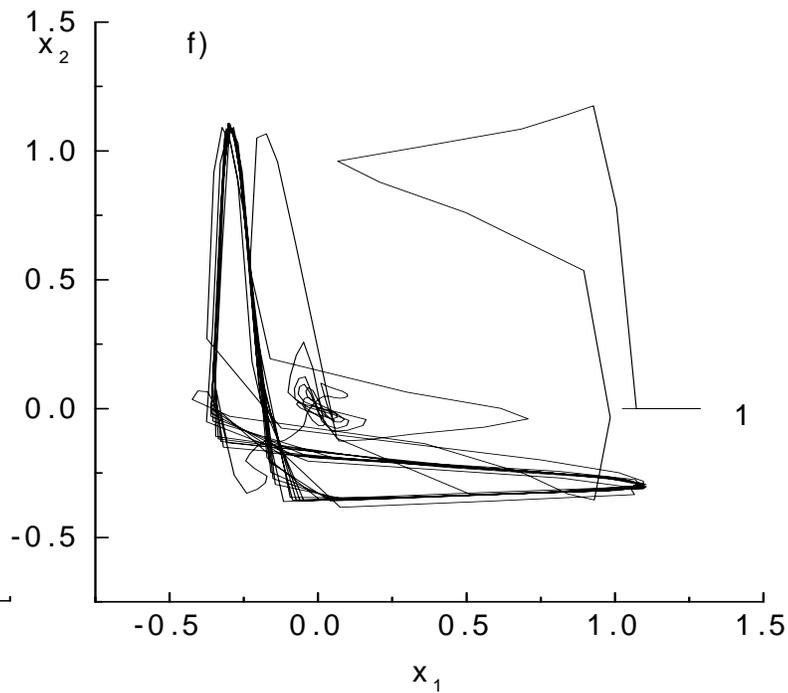
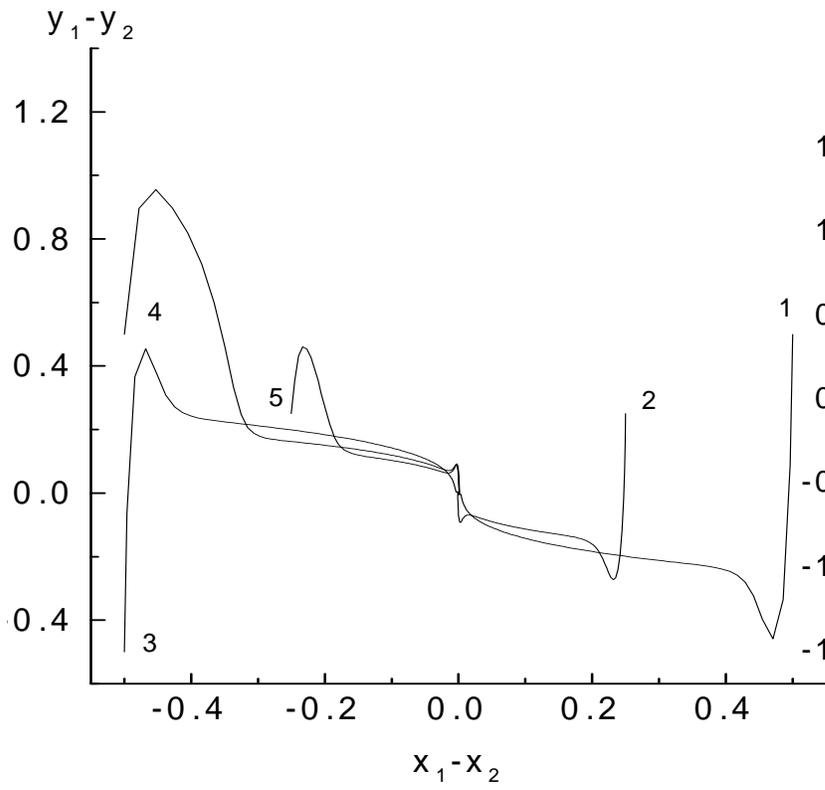
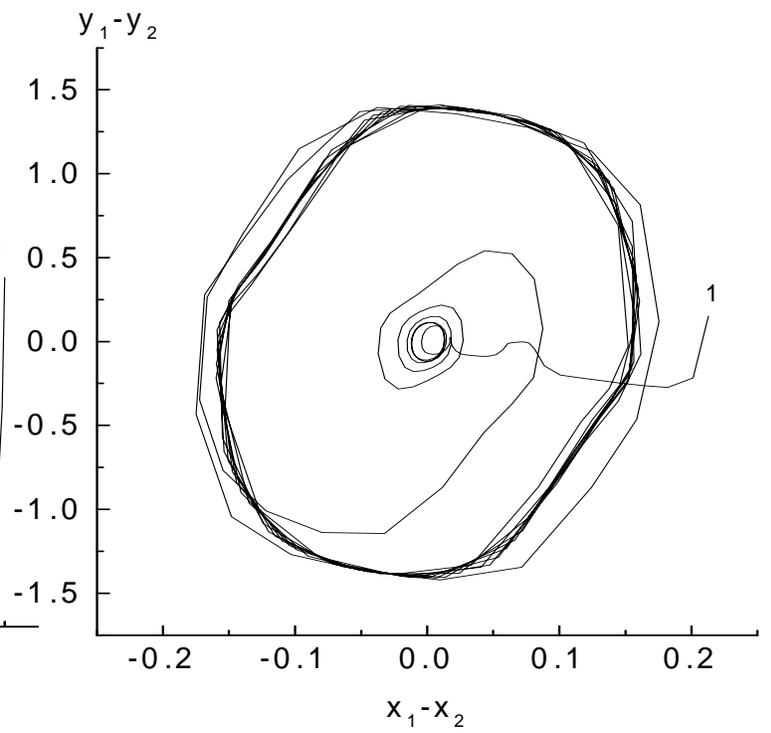
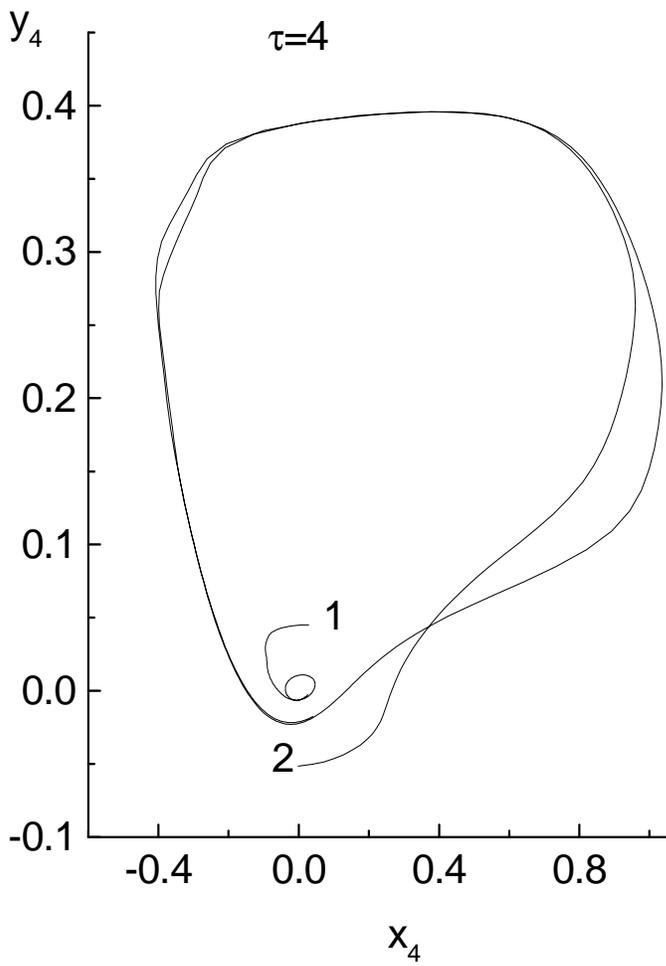
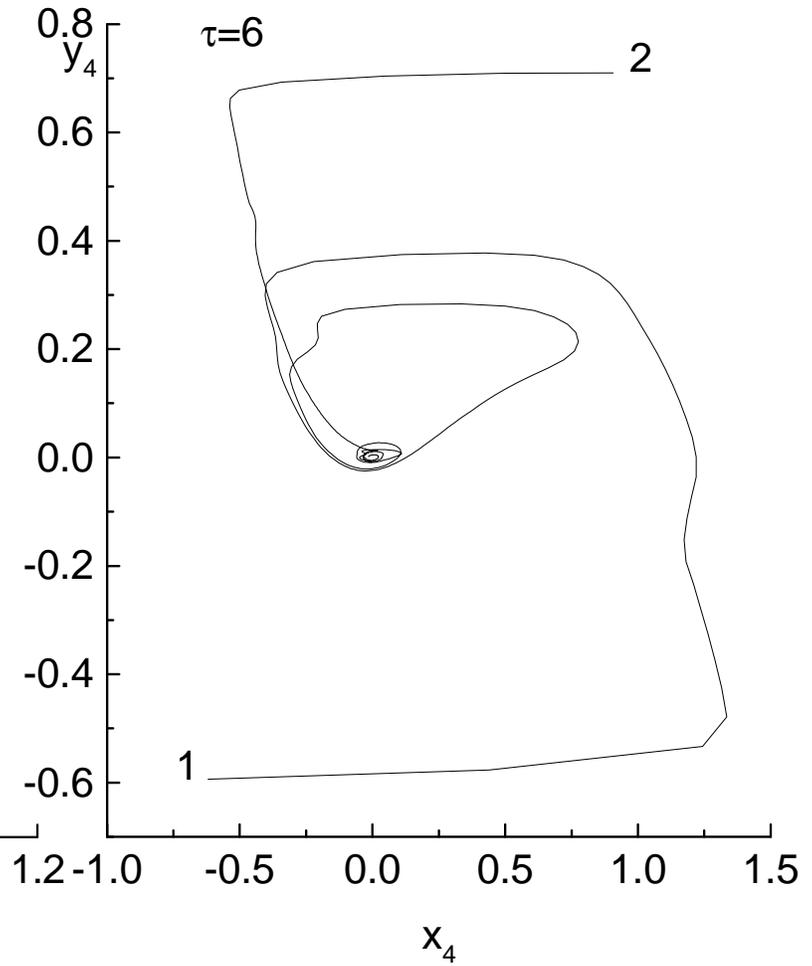
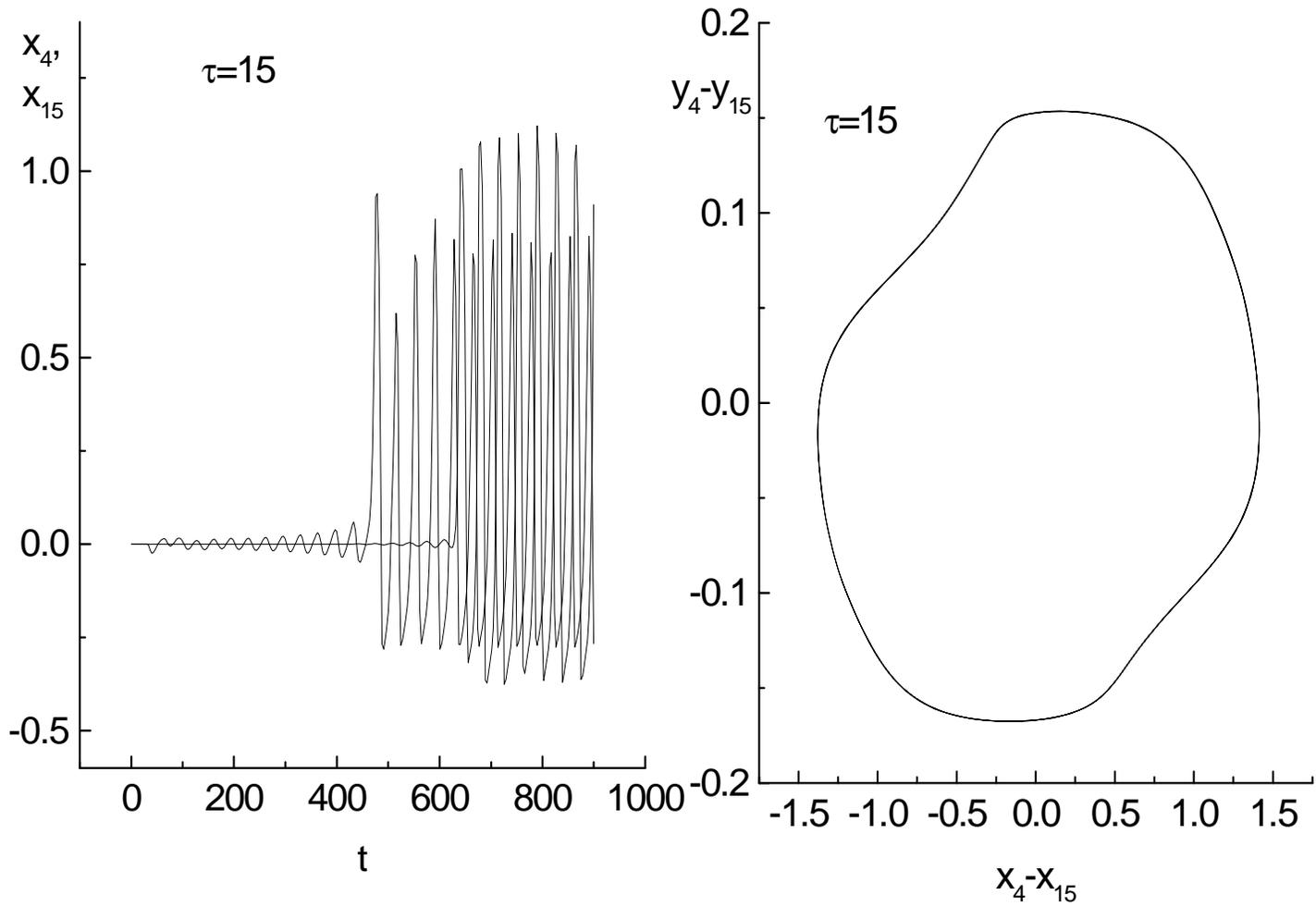
\end{document}